\documentclass[twocolumn]{aastex63}

\usepackage[flushleft]{threeparttable}
\usepackage{tabularx}
\usepackage{graphicx}
\usepackage{txfonts}
\usepackage{subfigure}

\received{June 1, 2019}
\revised{January 10, 2019}
\accepted{\today}
\submitjournal{AJ}

\shorttitle{WASP-150b \& WASP-176b}
\shortauthors{Benjamin F. Cooke et al.}

\graphicspath{{./}{figures_post-ref2/}}

\newcommand{\epochA}{$7217.2614^{+0.0004}_{-0.0004}$}
\newcommand{\periodA}{$5.644207^{+0.000003}_{-0.000004}$}
\newcommand{\deltafluxA}{$0.0044^{+0.0001}_{-0.0001}$}
\newcommand{\transitdurationA}{$0.1299^{+0.001}_{-0.0015}$}
\newcommand{\impactA}{$0.758^{+0.011}_{-0.014}$}
\newcommand{\eccentricityA}{$0.3775^{+0.0038}_{-0.0029}$}
\newcommand{\inclinationA}{$84.01^{+0.25}_{-0.2}$}
\newcommand{\stardensityA}{$0.439^{+0.032}_{-0.025}$}
\newcommand{\starteffA}{$6218.0^{+49.0}_{-45.0}$}
\newcommand{\starfehA}{$0.156^{+0.1}_{-0.089}$}
\newcommand{\starmassA}{$1.394^{+0.07}_{-0.049}$}
\newcommand{\starradiusA}{$1.651^{+0.024}_{-0.03}$}
\newcommand{\starloggA}{$4.147^{+0.027}_{-0.02}$}
\newcommand{\planetmassA}{$8.46^{+0.28}_{-0.2}$}
\newcommand{\planetradiusA}{$1.07^{+0.024}_{-0.025}$}
\newcommand{\planetdensityA}{$6.44^{+0.50}_{-0.47}$}
\newcommand{\planetloggA}{$4.263^{+0.024}_{-0.023}$}
\newcommand{\planettempA}{$1460.0^{+11.0}_{-11.0}$}
\newcommand{\planetsepA}{$0.0694^{+0.0011}_{-0.0008}$}

\newcommand{\starRAA}{17\,:\,37\,:\,03.14}
\newcommand{\starDecA}{+53\,:\,01\,:\,16.4}
\newcommand{\starmagVA}{$12.03$}
\newcommand{\starmagJA}{$11.06$}
\newcommand{\starmagGA}{$11.92$}
\newcommand{\starclassspecA}{F8}
\newcommand{\starteffspecA}{$6250\pm80$}
\newcommand{\starloggspecA}{$4.23\pm0.13$}
\newcommand{\starfehspecA}{$0.18\pm0.11$}
\newcommand{\starvsinispecA}{$8.82\pm0.95$}
\newcommand{\starparallaxgaiaA}{$1.865\pm0.020$}
\newcommand{\stardistgaiaA}{$536\pm6$}
\newcommand{\starpmragaiaA}{$-4.289\pm0.041$}
\newcommand{\starpmdecgaiaA}{$7.000\pm0.040$}
\newcommand{\starteffgaiaA}{$6093^{+188}_{-70}$}
\newcommand{\starradiusgaiaA}{$1.706^{+0.040}_{-0.101}$}
\newcommand{\starlumgaiaA}{$3.616\pm0.0.073$}
\newcommand{\starmassBAGEMASSA}{$1.346\pm0.029$}
\newcommand{\starageisoA}{$2.950\pm0.229$}

\newcommand{\starfehBAGEMASSA}{$0.204\pm0.079$}
\newcommand{\circtimeA}{$5$}
\newcommand{\epochB}{$8234.1771^{+0.0007}_{-0.0007}$}
\newcommand{\periodB}{$3.899052^{+0.000005}_{-0.000005}$}
\newcommand{\deltafluxB}{$0.0064^{+0.0002}_{-0.0002}$}
\newcommand{\transitdurationB}{$0.2147^{+0.0021}_{-0.0019}$}
\newcommand{\impactB}{$0.347^{+0.098}_{-0.12}$}
\newcommand{\eccentricityB}{0.0 (Fixed)}
\newcommand{\inclinationB}{$86.7^{+1.3}_{-1.1}$}
\newcommand{\stardensityB}{$0.263^{+0.03}_{-0.032}$}
\newcommand{\starteffB}{$5941.0^{+77.0}_{-79.0}$}
\newcommand{\starfehB}{$0.164^{+0.081}_{-0.082}$}
\newcommand{\starmassB}{$1.345^{+0.08}_{-0.13}$}
\newcommand{\starradiusB}{$1.925^{+0.047}_{-0.044}$}
\newcommand{\starloggB}{$3.995^{+0.037}_{-0.053}$}
\newcommand{\planetmassB}{$0.855^{+0.072}_{-0.069}$}
\newcommand{\planetradiusB}{$1.505^{+0.05}_{-0.045}$}
\newcommand{\planetdensityB}{$0.234^{+0.032}_{-0.032}$}
\newcommand{\planetloggB}{$2.972^{+0.047}_{-0.053}$}
\newcommand{\planettempB}{$1721.0^{+28.0}_{-21.0}$}
\newcommand{\planetsepB}{$0.0535^{+0.001}_{-0.0019}$}

\newcommand{\starRAB}{20\,:\,54\,:\,44.94}
\newcommand{\starDecB}{+09\,:\,10\,:\,44.5}
\newcommand{\starmagVB}{$12.01$}
\newcommand{\starmagJB}{$10.99$}
\newcommand{\starmagGB}{$11.94$}
\newcommand{\starclassspecB}{F9}
\newcommand{\starteffspecB}{$6100\pm100$}
\newcommand{\starloggspecB}{$4.0\pm0.2$}
\newcommand{\starfehspecB}{$0.15\pm0.08$}
\newcommand{\starvsinispecB}{$3.8\pm1.0$}
\newcommand{\starparallaxgaiaB}{$1.731\pm0.036$}
\newcommand{\stardistgaiaB}{$578\pm12$}
\newcommand{\starpmragaiaB}{$-6.192\pm0.064$}
\newcommand{\starpmdecgaiaB}{$-4.954\pm0.062$}
\newcommand{\starteffgaiaB}{$5902^{+20}_{-36}$}
\newcommand{\starradiusgaiaB}{$1.945^{+0.024}_{-0.013}$}
\newcommand{\starlumgaiaB}{$4.136\pm0.130$}
\newcommand{\starmassBAGEMASSB}{$1.270\pm0.025$}
\newcommand{\starageisoB}{$4.810\pm0.191$}

\newcommand{\starfehBAGEMASSB}{$0.215\pm0.069$}
\newcommand{\periodAshort}{$5.6$}
\newcommand{\eccentricityAshort}{$0.38$}
\newcommand{\starmassAshort}{$1.4$}
\newcommand{\starradiusAshort}{$1.7$}
\newcommand{\planetmassAshort}{$8.5$}
\newcommand{\planetradiusAshort}{$1.1$}
\newcommand{\planetdensityAshort}{$6.4$}

\newcommand{\periodBshort}{$3.9$}
\newcommand{\starmassBshort}{$1.3$}
\newcommand{\starradiusBshort}{$1.9$}
\newcommand{\planetradiusBshort}{$1.5$}
\newcommand{\planetmassBshort}{$0.86$}
\newcommand{\planetdensityBshort}{$0.23$}

\begin{document}

\title{Two transiting hot Jupiters from the WASP survey: WASP-150b and WASP-176b}

\correspondingauthor{Benjmain F. Cooke}
\email{b.cooke@warwick.ac.uk}

\author[0000-0002-8824-9956]{Benjamin F. Cooke}
\affiliation{Department of Physics, University of Warwick, Gibbet Hill Road, Coventry CV4 7AL, UK}
\affiliation{Centre for Exoplanets and Habitability, University of Warwick, Gibbet Hill Road, Coventry CV4 7AL, UK}

\author{Don Pollacco}
\affiliation{Department of Physics, University of Warwick, Gibbet Hill Road, Coventry CV4 7AL, UK}
\affiliation{Centre for Exoplanets and Habitability, University of Warwick, Gibbet Hill Road, Coventry CV4 7AL, UK}

\author{Y. Almleaky}
\affiliation{Space and Astronomy Department, Faculty of Science, King Abdulaziz University, 21589 Jeddah, Saudi Arabia}
\affiliation{King Abdullah Centre for Crescent Observations and Astronomy, Makkah Clock, Mecca 24231, Saudi Arabia}

\author{K. Barkaoui}
\affiliation{Astrobiology Research Unit, Universit\'e de Li\`ege, Belgium}
\affiliation{Oukaimeden Observatory, High Energy Physics and Astrophysics Laboratory, Cadi Ayyad University, Marrakech, Morocco}

\author{Z. Benkhaldoun}
\affiliation{Oukaimeden Observatory, High Energy Physics and Astrophysics Laboratory, Cadi Ayyad University, Marrakech, Morocco}

\author{James A. Blake}
\affiliation{Department of Physics, University of Warwick, Gibbet Hill Road, Coventry CV4 7AL, UK}
\affiliation{Centre for Exoplanets and Habitability, University of Warwick, Gibbet Hill Road, Coventry CV4 7AL, UK}

\author{Fran{\c c}ois Bouchy}
\affiliation{Observatoire astronomique de l'Universit\'e de Geneve, 51 ch. des Maillettes, 1290 Sauverny, Switzerland}

\author{Panos Boumis}
\affiliation{Institute for Astronomy, Astrophysics, Space Applications and Remote Sensing, National Observatory of Athens, 15236 Penteli, Greece}

\author{D. J. A. Brown}
\affiliation{Department of Physics, University of Warwick, Gibbet Hill Road, Coventry CV4 7AL, UK}
\affiliation{Centre for Exoplanets and Habitability, University of Warwick, Gibbet Hill Road, Coventry CV4 7AL, UK}

\author{Ivan Bruni}
\affiliation{INAF -- Osservatorio Astronomico di Bologna, Via Ranzani 1, I--40127 Bologna, Italy}

\author{A. Burdanov}
\affiliation{Astrobiology Research Unit, Universit\'e de Li\`ege, Belgium}

\author{Andrew Collier Cameron}
\affiliation{Centre for Exoplanet Science, SUPA, School of Physics and Astronomy, University of St Andrews, North Haugh, St Andrews KY16 9SS, UK}

\author{Paul Chote}
\affiliation{Department of Physics, University of Warwick, Gibbet Hill Road, Coventry CV4 7AL, UK}
\affiliation{Centre for Exoplanets and Habitability, University of Warwick, Gibbet Hill Road, Coventry CV4 7AL, UK}

\author{A. Daassou}
\affiliation{Oukaimeden Observatory, High Energy Physics and Astrophysics Laboratory, Cadi Ayyad University, Marrakech, Morocco}

\author{Giuseppe D'ago}
\affiliation{Instituto de Astrof\'{i}sica, Facultad de F\'{i}sica, Pontificia Universidad Cat\'{o}lica de Chile, Av. Vicu\~{n}a Mackenna 4860, 7820436 Macul, Santiago, Chile}

\author{Shweta Dalal}
\affiliation{Institut d’Astrophysique de Paris, UMR7095 CNRS, Universite Pierre \& Marie Curie, 98bis boulevard Arago, 75014 Paris, France}

\author{Mario Damasso}
\affiliation{INAF -- Osservatorio Astrofisico di Torino, via Osservatorio 20, I-10025, Pino Torinese, Italy}

\author{L. Delrez}
\affiliation{Cavendish  Laboratory, J J Thomson Avenue, Cambridge CB3 0HE, UK}

\author{A. P. Doyle}
\affiliation{Department of Physics, University of Warwick, Gibbet Hill Road, Coventry CV4 7AL, UK}

\author{E. Ducrot}
\affiliation{Astrobiology Research Unit, Universit\'e de Li\`ege, Belgium}

\author{M. Gillon}
\affiliation{Astrobiology Research Unit, Universit\'e de Li\`ege, Belgium}

\author{G. H{\'e}brard}
\affiliation{Institut d’Astrophysique de Paris, UMR7095 CNRS, Universite Pierre \& Marie Curie, 98bis boulevard Arago, 75014 Paris, France}

\author{C. Hellier}
\affiliation{INAF -- Osservatorio Astrofisico di Torino, via Osservatorio 20, I-10025, Pino Torinese, Italy}

\author{Thomas Henning}
\affiliation{Max Planck Institute for Astronomy, K\"{o}nigstuhl 17, D-69117, Heidelberg, Germany}

\author{E. Jehin}
\affiliation{Space sciences, Technologies and Astrophysics Research (STAR) Institute, Universit\'e de Li\`ege, Belgium}

\author{Flavien Kiefer}
\affiliation{Institut d’Astrophysique de Paris, UMR7095 CNRS, Universite Pierre \& Marie Curie, 98bis boulevard Arago, 75014 Paris, France}

\author{George W. King}
\affiliation{Department of Physics, University of Warwick, Gibbet Hill Road, Coventry CV4 7AL, UK}
\affiliation{Centre for Exoplanets and Habitability, University of Warwick, Gibbet Hill Road, Coventry CV4 7AL, UK}

\author{Alexios Liakos}
\affiliation{Institute for Astronomy, Astrophysics, Space Applications and Remote Sensing, National Observatory of Athens, 15236 Penteli, Greece}

\author{Th{\'e}o Lopez}
\affiliation{Aix Marseille Univ, CNRS, CNES, LAM, Marseille, France}

\author{Luigi Mancini}
\affiliation{Department of Physics, University of Rome Tor Vergata, Via della Ricerca Scientifica 1, I-00133, Rome, Italy}
\affiliation{Max Planck Institute for Astronomy, K\"{o}nigstuhl 17, D-69117, Heidelberg, Germany}
\affiliation{INAF -- Osservatorio Astrofisico di Torino, via Osservatorio 20, I-10025, Pino Torinese, Italy}
\affiliation{International Institute for Advanced Scientific Studies (IIASS), Via G. Pellegrino 19, I-84019, Vietri sul Mare (SA), Italy}

\author{Rosemary Mardling}
\affiliation{School of Physics and Astronomy, Monash University, Victoria, 3800, Australia}
\affiliation{Observatoire astronomique de l'Universit\'e de Geneve, 51 ch. des Maillettes, 1290 Sauverny, Switzerland}

\author{P. F. L. Maxted}
\affiliation{Astrophysics Group, Lennard-Jones Laboratories, Keele University, Staffordshire ST5 5BG, UK}

\author{James McCormac}
\affiliation{Department of Physics, University of Warwick, Gibbet Hill Road, Coventry CV4 7AL, UK}
\affiliation{Centre for Exoplanets and Habitability, University of Warwick, Gibbet Hill Road, Coventry CV4 7AL, UK}

\author{C. Murray}
\affiliation{Cavendish  Laboratory, J J Thomson Avenue, Cambridge CB3 0HE, UK}

\author{Louise D. Nielsen}
\affiliation{Observatoire astronomique de l'Universit\'e de Geneve, 51 ch. des Maillettes, 1290 Sauverny, Switzerland}

\author{Hugh Osborn}
\affiliation{Aix Marseille Univ, CNRS, CNES, LAM, Marseille, France}
\affiliation{Center for Space and Habitability, University of Bern, Gesellschaftsstrasse 6, 3012 Bern, Switzerland}

\author{E. Palle}
\affiliation{Instituto de Astrof\'\i sica de Canarias (IAC), 38205 La Laguna, Tenerife, Spain}
\affiliation{Departamento de Astrof\'\i sica, Universidad de La Laguna (ULL), 38206 La Laguna, Tenerife, Spain}

\author{Francesco Pepe}
\affiliation{Observatoire astronomique de l'Universit\'e de Geneve, 51 ch. des Maillettes, 1290 Sauverny, Switzerland}

\author{F. J. Pozuelos}
\affiliation{Astrobiology Research Unit, Universit\'e de Li\`ege, Belgium}
\affiliation{Space sciences, Technologies and Astrophysics Research (STAR) Institute, Universit\'e de Li\`ege, Belgium}

\author{J. Prieto-Arranz}
\affiliation{Instituto de Astrof\'\i sica de Canarias (IAC), 38205 La Laguna, Tenerife, Spain}
\affiliation{Departamento de Astrof\'\i sica, Universidad de La Laguna (ULL), 38206 La Laguna, Tenerife, Spain}

\author{D. Queloz}
\affiliation{Observatoire astronomique de l'Universit\'e de Geneve, 51 ch. des Maillettes, 1290 Sauverny, Switzerland}
\affiliation{Cavendish  Laboratory, J J Thomson Avenue, Cambridge CB3 0HE, UK}

\author{Nicole Schanche}
\affiliation{Centre for Exoplanet Science, SUPA, School of Physics and Astronomy, University of St Andrews, North Haugh, St Andrews KY16 9SS, UK}

\author{Damien S{\'e}gransan}
\affiliation{Observatoire astronomique de l'Universit\'e de Geneve, 51 ch. des Maillettes, 1290 Sauverny, Switzerland}

\author{Barry Smalley}
\affiliation{Astrophysics Group, Lennard-Jones Laboratories, Keele University, Staffordshire ST5 5BG, UK}

\author{John Southworth}
\affiliation{Astrophysics Group, Lennard-Jones Laboratories, Keele University, Staffordshire ST5 5BG, UK}

\author{S. Thompson}
\affiliation{Cavendish  Laboratory, J J Thomson Avenue, Cambridge CB3 0HE, UK}

\author{Oliver Turner}
\affiliation{Observatoire astronomique de l'Universit\'e de Geneve, 51 ch. des Maillettes, 1290 Sauverny, Switzerland}

\author{St{\'e}phane Udry}
\affiliation{Observatoire astronomique de l'Universit\'e de Geneve, 51 ch. des Maillettes, 1290 Sauverny, Switzerland}

\author{S. Velasco}
\affiliation{Instituto de Astrof\'\i sica de Canarias (IAC), 38205 La Laguna, Tenerife, Spain}
\affiliation{Departamento de Astrof\'\i sica, Universidad de La Laguna (ULL), 38206 La Laguna, Tenerife, Spain}

\author{Richard West}
\affiliation{Department of Physics, University of Warwick, Gibbet Hill Road, Coventry CV4 7AL, UK}
\affiliation{Centre for Exoplanets and Habitability, University of Warwick, Gibbet Hill Road, Coventry CV4 7AL, UK}

\author{Pete Wheatley}
\affiliation{Department of Physics, University of Warwick, Gibbet Hill Road, Coventry CV4 7AL, UK}
\affiliation{Centre for Exoplanets and Habitability, University of Warwick, Gibbet Hill Road, Coventry CV4 7AL, UK}

\author{John Alikakos}
\affiliation{Institute for Astronomy, Astrophysics, Space Applications and Remote Sensing, National Observatory of Athens, 15236 Penteli, Greece}



\begin{abstract}

\noindent
We report the discovery of two transiting exoplanets from the WASP survey, WASP-150b and WASP-176b.
\newline
WASP-150b is an eccentric ($e$ = \eccentricityAshort) hot Jupiter on a \periodAshort\,day orbit around a $V$ = \starmagVA, \starclassspecA~main-sequence host. The host star has a mass and radius of \starmassAshort\,$\rm M_{\odot}$ and \starradiusAshort\,$\rm R_{\odot}$ respectively. WASP-150b has a mass and radius of \planetmassAshort\,$\rm M_J$ and \planetradiusAshort\,$\rm R_J$, leading to a large planetary bulk density of \planetdensityAshort\,$\rm \rho_J$. WASP-150b is found to be $\sim3$\,Gyr old, well below its circularisation timescale, supporting the eccentric nature of the planet.
\newline
WASP-176b is a hot Jupiter planet on a \periodBshort\,day orbit around a $V$ = \starmagVB, \starclassspecB~sub-giant host. The host star has a mass and radius of \starmassBshort\,$\rm M_{\odot}$ and \starradiusBshort\,$\rm R_{\odot}$. WASP-176b has a mass and radius of \planetmassBshort\,$\rm M_J$ and \planetradiusBshort\,$\rm R_J$ respectively, leading to a planetary bulk density of \planetdensityBshort\,$\rm \rho_J$.

\end{abstract}

\keywords{Planetary systems -–- Stars: individual: WASP-150, WASP-176 –-- Techniques: photometric, radial velocities}


\section{Introduction}
\label{sec:Introduction}

As of October 2019, over 4000 exoplanets have been  verified\footnote{\url{https://exoplanetarchive.ipac.caltech.edu/}}. Of these planets, over 3000 have been discovered using the transit method. These results have shown that exoplanet populations are both very common and very diverse \citep{Batalha2014}, with a wide range of system parameters found thus far. The transit discoveries have been made using both space-based surveys, for example Kepler/K2 \citep{Borucki2010,Howell2014}, and ground-based surveys, including WASP \citep{Pollacco2006}, HATNet/HATSouth \citep{Bakos2018}, KELT \citep{Pepper2007}, TRAPPIST \citep{Jehin2011} and NGTS \citep{Wheatley2017}. This field is expected to be expanded upon even further in the coming years with the yield of TESS \citep{Ricker2015} and other future planned missions such as JWST \citep{Gardner2006}, PLATO \citep{Rauer2016} and ARIEL \citep{Pascale2018}.

Within this sample of exoplanets lies the region corresponding to hot Jupiters. A type of planet noticeably absent from our own solar system, hot Jupiters are planets with masses comparable to Jupiter but with orbital periods on the order of days. Current estimates put the occurrence rate of giant planets within 5\,-\,10 au around FGK stars at 10\,-\,20\% \citep{Cumming2008,Mayor2011}. 
Further detections and precise characterisations of hot Jupiter exoplanets will improve our ability to study this population using statistical methods, and hence allow for a better understanding of this unique type of planet.



Exoplanets are broadly characterised using a combination of the transit method \citep{Henry2000,Charbonneau2000,2010exop.book...55W} and radial velocity measurements. When combined these methods allow for the measurement of radius, mass and density which can lead to inferences of composition.

This paper discusses the discovery and characterisation of the transiting hot Jupiters WASP-150b and WASP-176b. Sections \ref{sec:SuperWASP discovery photometry}, \ref{sec:Spectroscopic follow-up} and \ref{sec:Photometric follow-up} detail the WASP discovery, spectroscopic follow-up and photometric follow-up, respectively. Section \ref{sec:imaging} discusses the high-spatial-resolution follow-up of WASP-150. Section \ref{sec:Results} outlines the analysis and derived parameters. Finally, section \ref{sec:Discussion and conclusions} summarises the discovery findings.

\section{SuperWASP discovery photometry}
\label{sec:SuperWASP discovery photometry}

The WASP project (now decommissioned) was split into north and south facilities with telescopes located at the Isaac Newton Group (ING) at the Observatorio del Roque de los Muchachos, La Palma, Spain and at the South African Astronomical Observatory (SAAO), Sutherland, RSA respectively. Both facilities consisted of 8 Canon 200\,mm f/1.8 lenses, each linked to an Andor e2v $2048\times2048$ pixel CCD. Each camera had a total field of view of $7.8^{\circ}\times7.8^{\circ}$ with a pixel scale of $13.7^{\prime\prime}$ \citep{Pollacco2006}.

\begin{figure}[htp]
    \begin{subfigure}[Folded photometry
    \label{fig:150_wasp_fold}]
    {\includegraphics[width=\columnwidth]{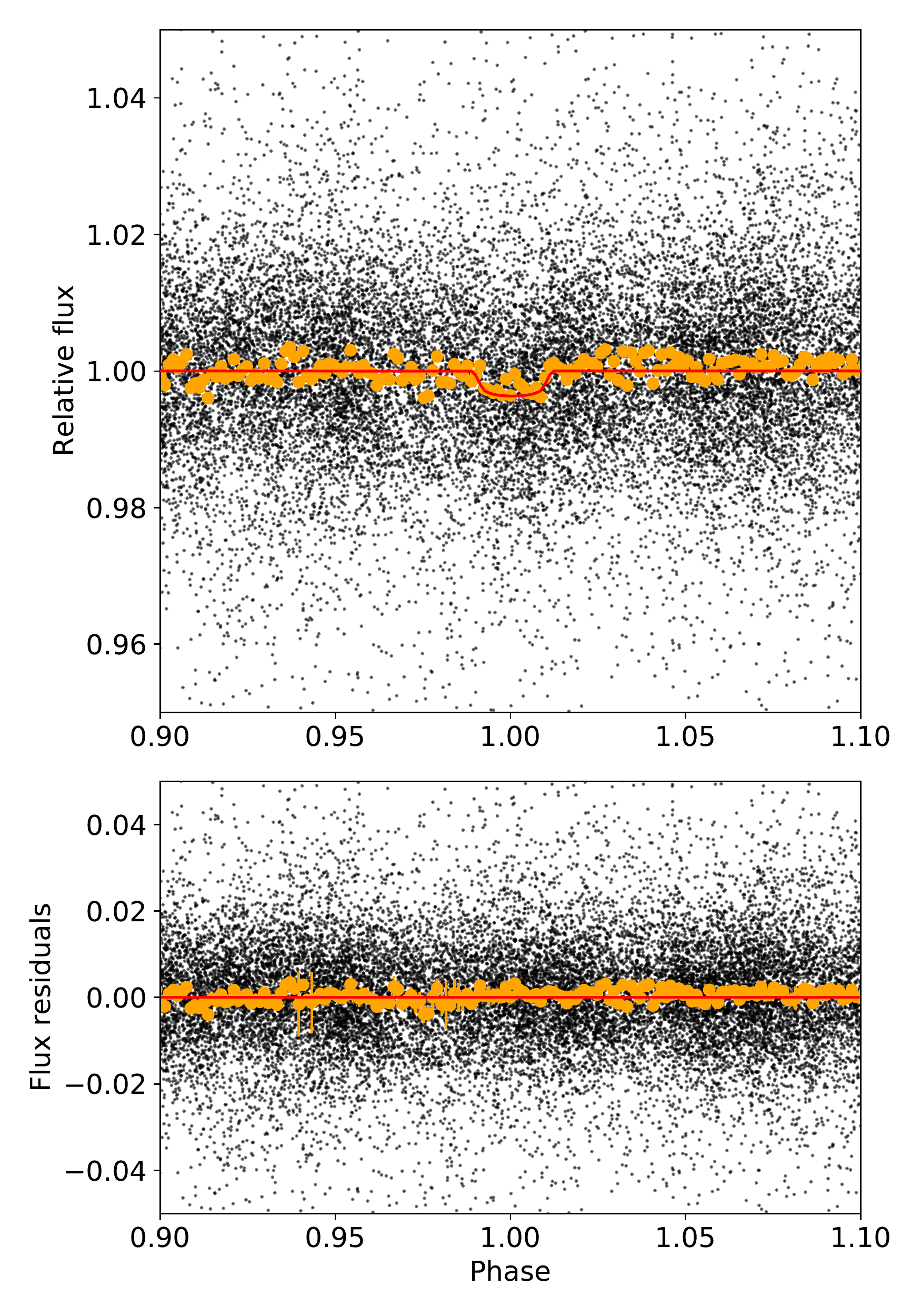}}
    \end{subfigure}
    \begin{subfigure}[BLS periodogram
    \label{fig:150_wasp_pgram}]
    {\includegraphics[width=\columnwidth]{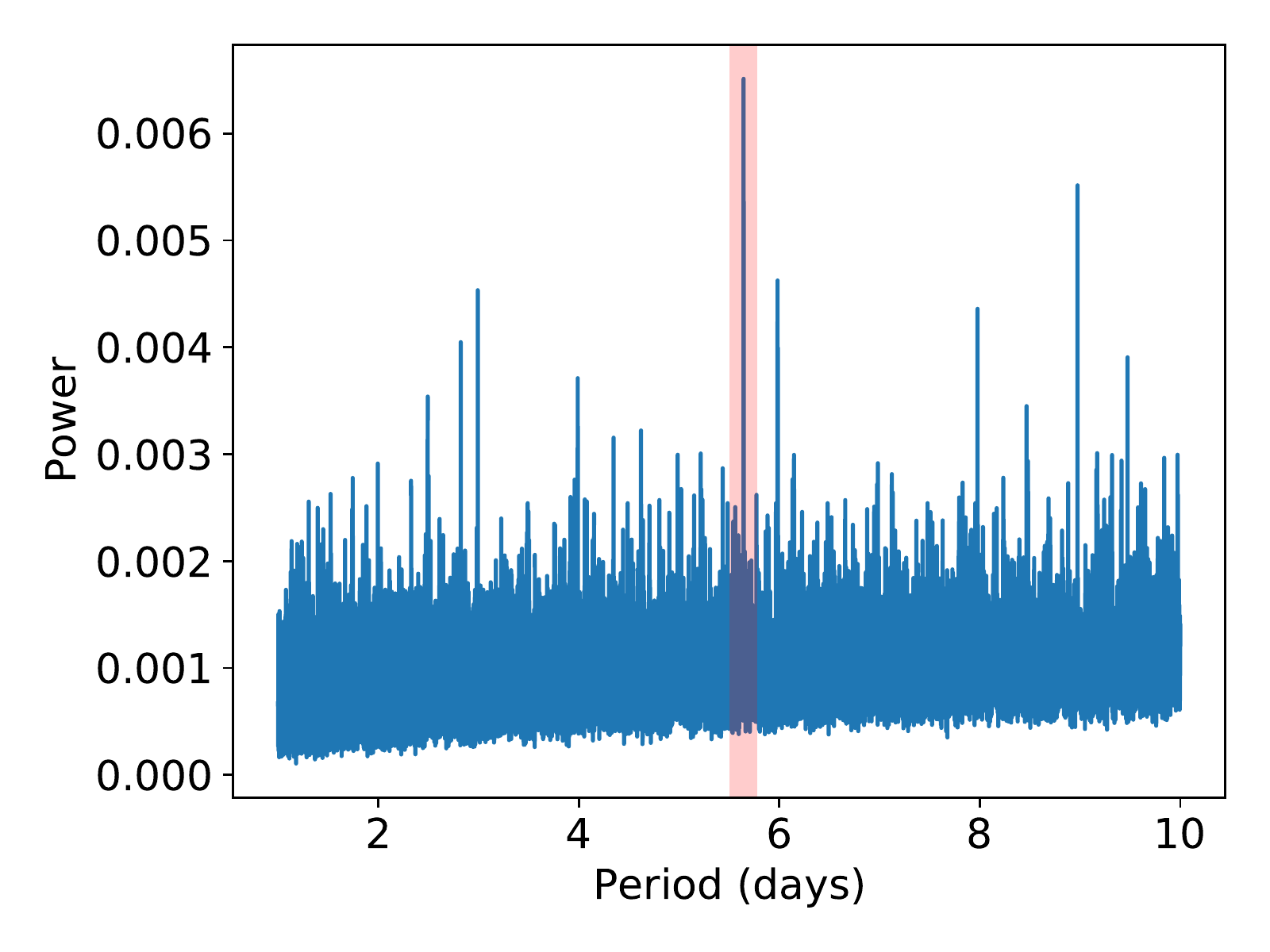}}
    \end{subfigure}
    \caption{(a) \textit{Upper panel}: Phase folded WASP data for WASP-150b binned to 10\,minutes. Data are shown in orange with the best fit model derived from MCMC analysis shown in red.
    \newline
    \textit{Lower panel}: Residuals from MCMC fit.
    \newline
    (b) BLS periodogram of WASP photometry. The period from our MCMC analysis is highlighted.}
\label{fig:150_wasp_phot}
\end{figure}

For WASP-150, a total of 99\,892 photometric data points were taken between 14 May 2004 and 4 August 2011. These data ranged across 38 transits. WASP-150b was flagged as a high priority candidate on 17 February 2014 and confirmed as a planet on 25 June 2015. Figure \ref{fig:150_wasp_phot} shows the WASP discovery curve folded by the best fit period and binned to 10\,minutes. Additionally we show the best fit transit model from our MCMC.


\begin{figure}[htp]
    \begin{subfigure}[Folded photometry
    \label{fig:176_wasp_fold}]
    {\includegraphics[width=\columnwidth]{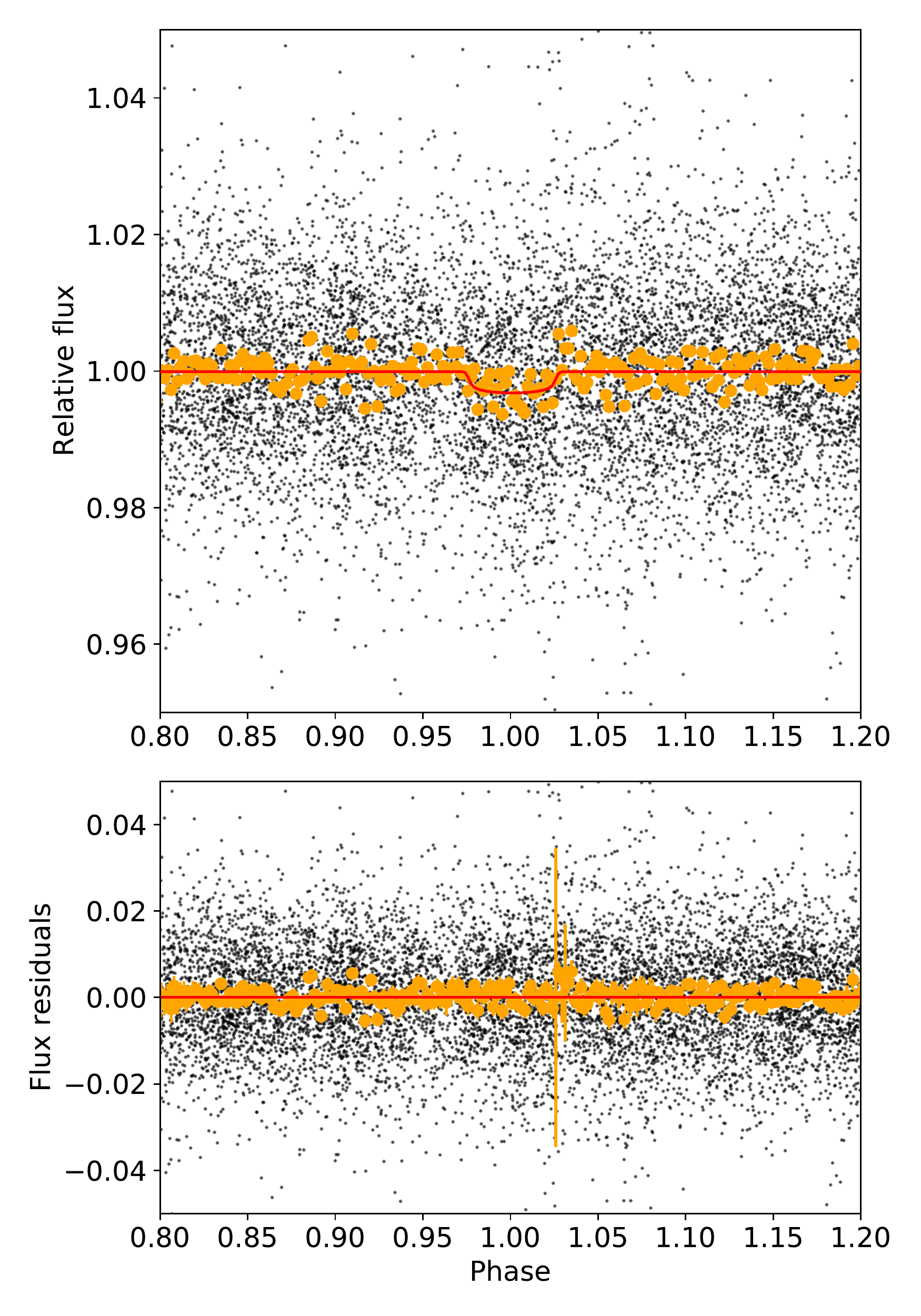}}
    \end{subfigure}
    \begin{subfigure}[BLS periodogram
    \label{fig:176_wasp_pgram}]
    {\includegraphics[width=\columnwidth]{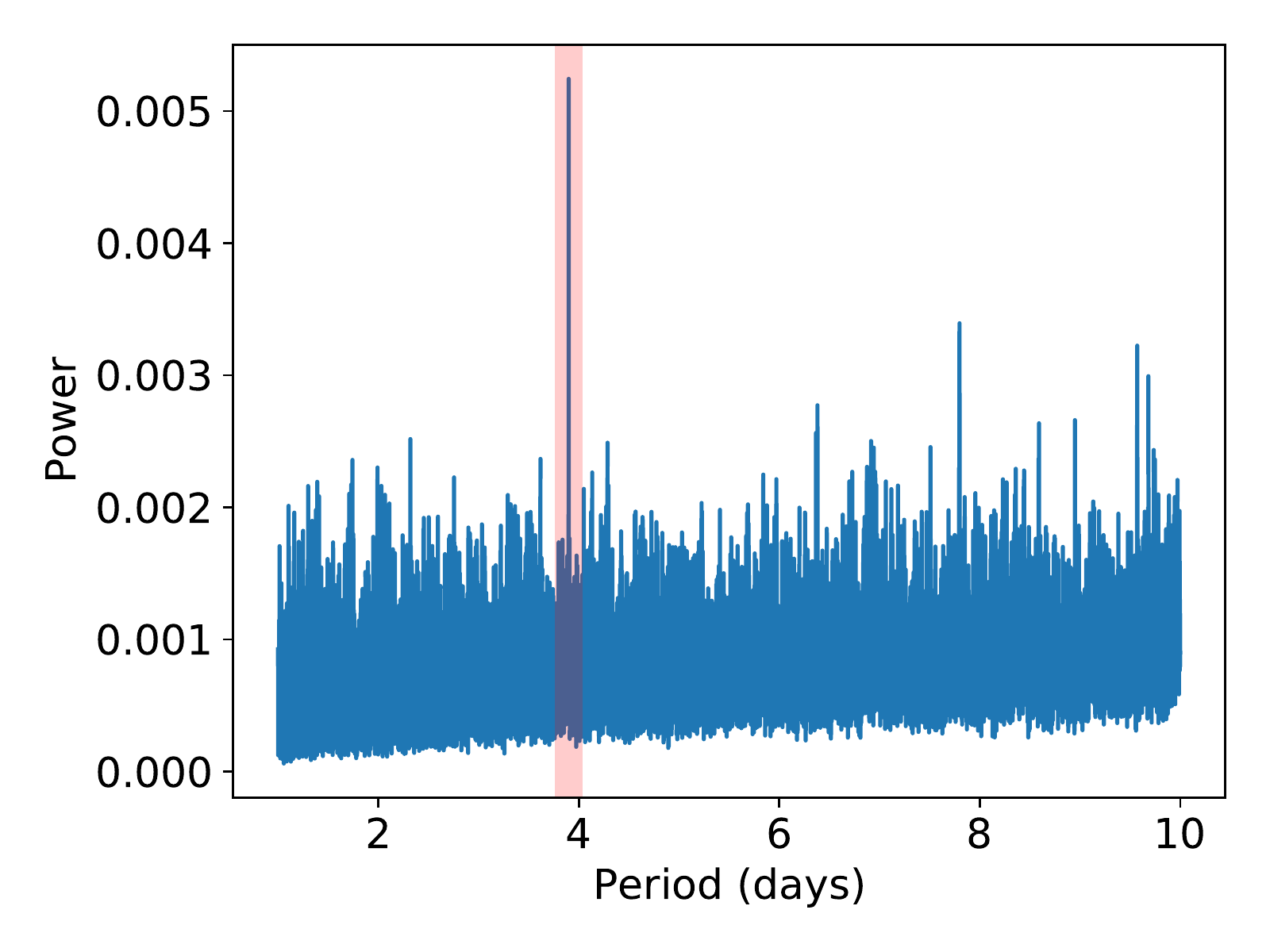}}
    \end{subfigure}
    \caption{(a) \textit{Upper panel}: Phase folded WASP data for WASP-176b binned to 10\,minutes. Data are shown in orange with the best fit model derived from MCMC analysis shown in red.
    \newline
    \textit{Lower panel}: Residuals from MCMC fit.
    \newline
    (b) BLS periodogram of WASP photometry. The period from our MCMC analysis is highlighted.}
\label{fig:176_wasp_phot}
\end{figure}

For WASP-176, a total of 23\,082 photometric data points were taken between 26 May 2004 and 6 October 2010. These data ranged across 30 transits. WASP-176b was flagged as a high priority candidate on 12 February 2014 and confirmed as a planet on 23 January 2018. Figure \ref{fig:176_wasp_phot} shows the WASP discovery curve folded by the best fit period and binned to 10\,minutes. Additionally we show the best fit transit model from our MCMC.


The SuperWASP data were reduced using the standard SuperWASP pipeline as described in \citet{Pollacco2006}. Analysis of the light curve was then carried out using the box least-squares (BLS) fit method, as in \citet{Kovacs2002}, and the SysRem detrending algorithm, described in \citet{Tamuz2005}. The results of BLS searches on the detrended data are shown in Figures \ref{fig:150_wasp_phot} and \ref{fig:176_wasp_phot}. The data were searched with a transit-search algorithm \citep{CollierCameron2007} and flagged as belonging to a planetary candidate. System parameters were then estimated from catalogue data and a Monte Carlo simulation \citep{CollierCameron2006}. These initial estimates produced a period of 5.644 days, a depth of 3.2\,mmag and a width of 2.4106 hours for WASP-150b and a period of 3.899 days, a depth of 3.4\,mmag and a width of 4.5292 hours for WASP-176b. 
Additional follow-up spectroscopy and photometry were then obtained to confirm and characterise the planets.

\begin{table}
	\centering
	\caption{Photometric properties}
	\label{tab:Photometric properties}
	\begin{tabular}{ccc}
		\hline
		\hline
		Parameter & WASP-150 & WASP-176\\
		\hline
        R.A. & \starRAA & \starRAB\\
        Dec. & \starDecA & \starDecB\\
        $V$ & \starmagVA & \starmagVB\\
        $J$ & \starmagJA & \starmagJB\\
        $G$ & \starmagGA & \starmagGB\\
		\hline
	\end{tabular}
\end{table}


\section{Spectroscopic follow-up}
\label{sec:Spectroscopic follow-up}

\subsection{WASP-150b}

WASP-150 was observed with the SOPHIE spectrograph, first to establish the planetary nature of the transiting candidate, then to characterize the secured planet by measuring in particular its mass and orbital eccentricity. SOPHIE is dedicated to high-precision RV  measurements at the 1.93\,m telescope of the Haute-Provence Observatory \citep{Perruchot2008,Bouchy2009b} and is widely used for WASP follow-up \citep[e.g.][]{CollierCameron2007b,Hebrard2013,Schanche2019}. We used its High-Efficiency mode with a resolving power $R=40\,000$ and slow readout mode. We obtained 22 observations between May 2014 and April 2015. Depending on weather conditions,  exposure times ranged between 8 and 33~minutes in order to maintain a signal-to-noise ratio as constant as possible among observations.

The spectra were extracted using the SOPHIE pipeline \citep{Bouchy2009b}, and the radial velocities were measured through weighted cross-correlation with a numerical mask \citep{Baranne1996,Pepe2002}. They were corrected for the CCD charge transfer inefficiency \citep{Bouchy2009b}, and their error bars were computed from the cross-correlation function (CCF) using the method presented by \citet{Boisse2010}. The monitoring of constant stars revealed no significant instrumental drifts during the epochs of observation, and none of the spectra were significantly affected by any moonlight or other sky background pollution.

The resulting CCFs have full width at half maximum (FWHM) of $14.1 \pm 0.1$\,km/s, and contrast that represents $\sim14$\,\% of the continuum. The lines are slightly broader than what is usually measured in High-Efficiency mode due to the stellar rotation of WASP-150 (we measured a projected rotational velocity $v$\,sin\,$i_*=8.3 \pm 1.0 $\,km/s from the parameters of the CCF using the calibration of \citet{Boisse2010}).

\begin{figure}[htp]
	\includegraphics[width=\columnwidth]{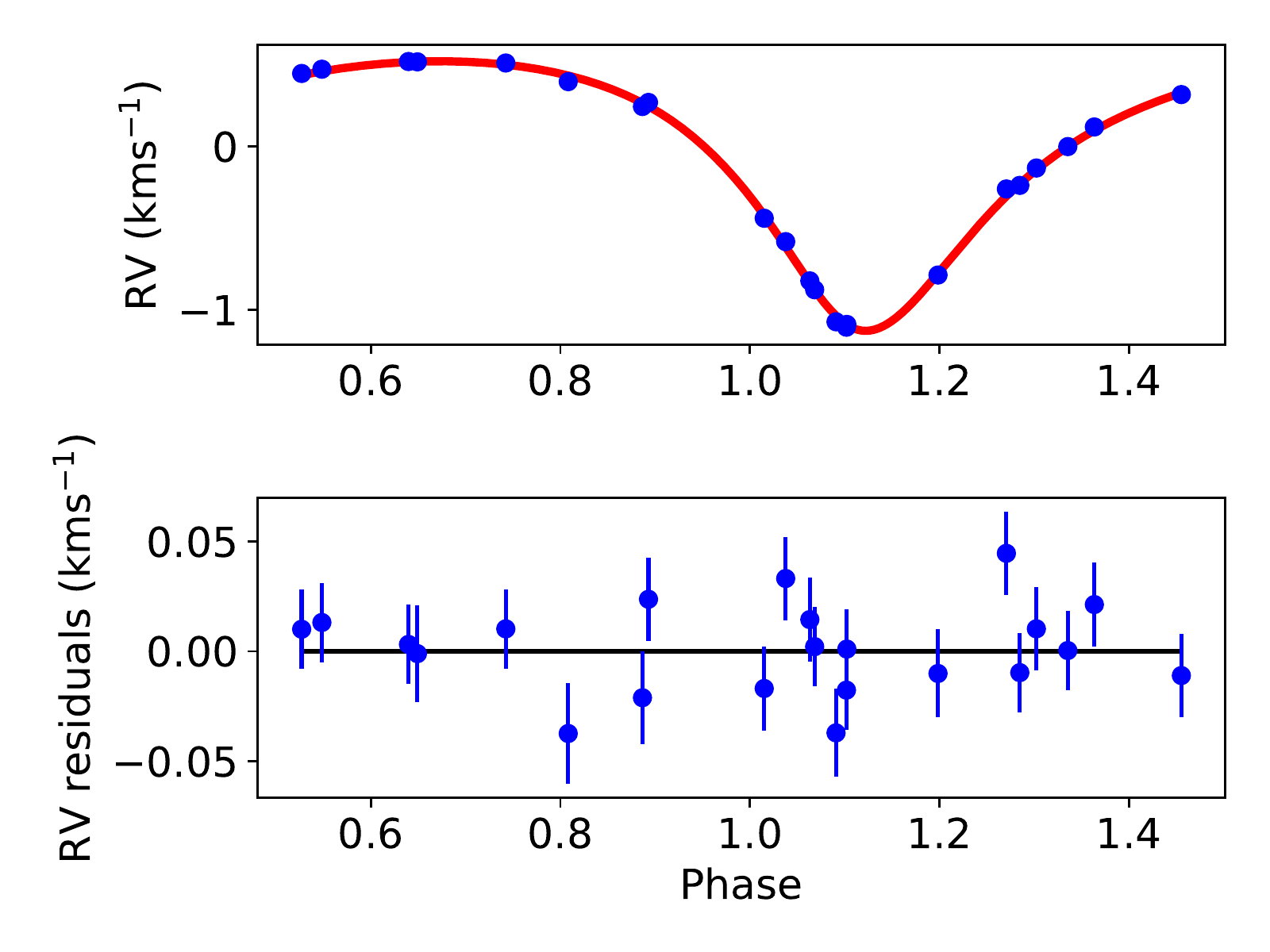}
    \caption{\textit{Upper panel}: Phase folded SOPHIE radial velocity observations of WASP-150. Data are shown in blue with the best fit model derived from MCMC analysis shown in red.
    \newline
    \textit{Lower panel}: Residuals from RV fit.}
    \label{fig:150_rv}
\end{figure}

The RVs have uncertainties around 19\,m/s. They show large variations in phase with the SuperWASP transit ephemeris for an eccentric orbit, and a semi-amplitudes of the order of 800\,m/s. This would correspond to a companion in the massive-planet regime. The SOPHIE RVs are shown in Figure \ref{fig:150_rv} with the best fit model and residuals from our MCMC analysis (see section \ref{sec:EXOFASTv2 analysis}). Data files can be found at \href{https://github.com/BenCooke95/W150-W176}{https://github.com/BenCooke95/W150-W176}.

\begin{figure}[htp]
	\includegraphics[width=\columnwidth]{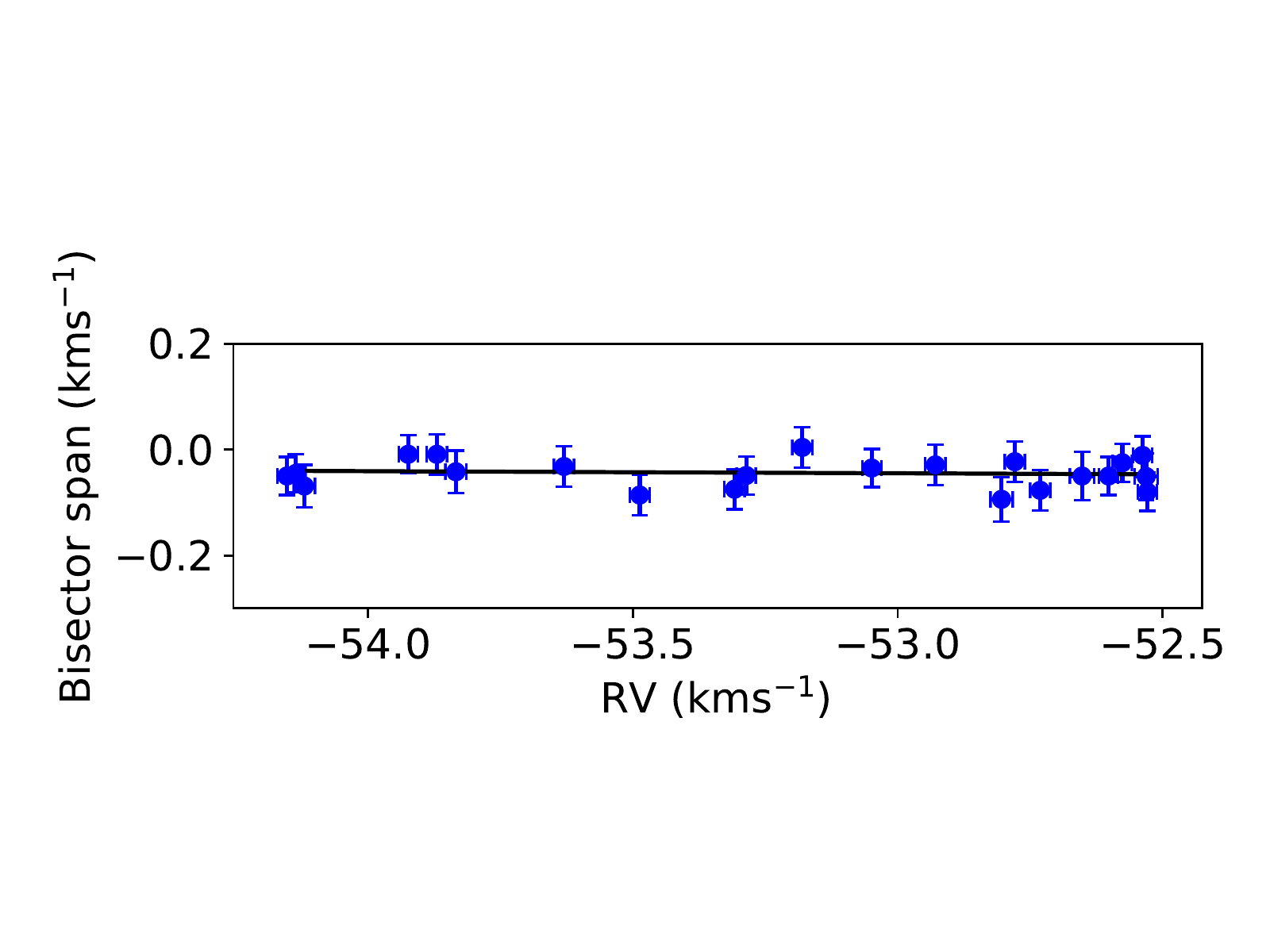}
    \caption{SOPHIE RV bisector span as a function of radial velocity for WASP-150. The solid black line shows the best weighted linear fit to the data. The lack of any significant gradient supports the assumption that the RV signal is produced by a planetary companion. The aspect ratio of the two axes is unity.}
    \label{fig:150_bis_span}
\end{figure}

Radial velocities measured using different stellar masks (F0, G2, K0, or K5) produce variations with similar amplitudes, so it is unlikely that these variations are produced by blend scenarios composed of stars of different spectral types. Similarly, the measured CCF bisector spans quantify possible shape variations of the spectral lines. They show a low dispersion of 27\,m/s, which agrees with their expected accuracy and is tiny by comparison to the large RV variations. We can thus conclude that the RV variations are not due to spectral-line profile changes attributable to blends or stellar activity, but rather to Doppler shifts due to a massive, eccentric planetary companion. Figure \ref{fig:150_bis_span} shows the bisector spans.

\subsection{WASP-176b}

High resolution spectroscopy for WASP-176 was performed with the CORALIE spectrograph \citep{Queloz2000,Pepe2017} on the Swiss 1.2\,m telescope at La Silla Observatory (Chile). In total we obtained 26 
measurements between July 2014 and July 2018. 
RVs were computed with the standard CORALIE data reduction pipeline by cross-correlating the spectra with a binary G2 mask \citep{Pepe2002}.

\begin{figure}[htp]
	\includegraphics[width=\columnwidth]{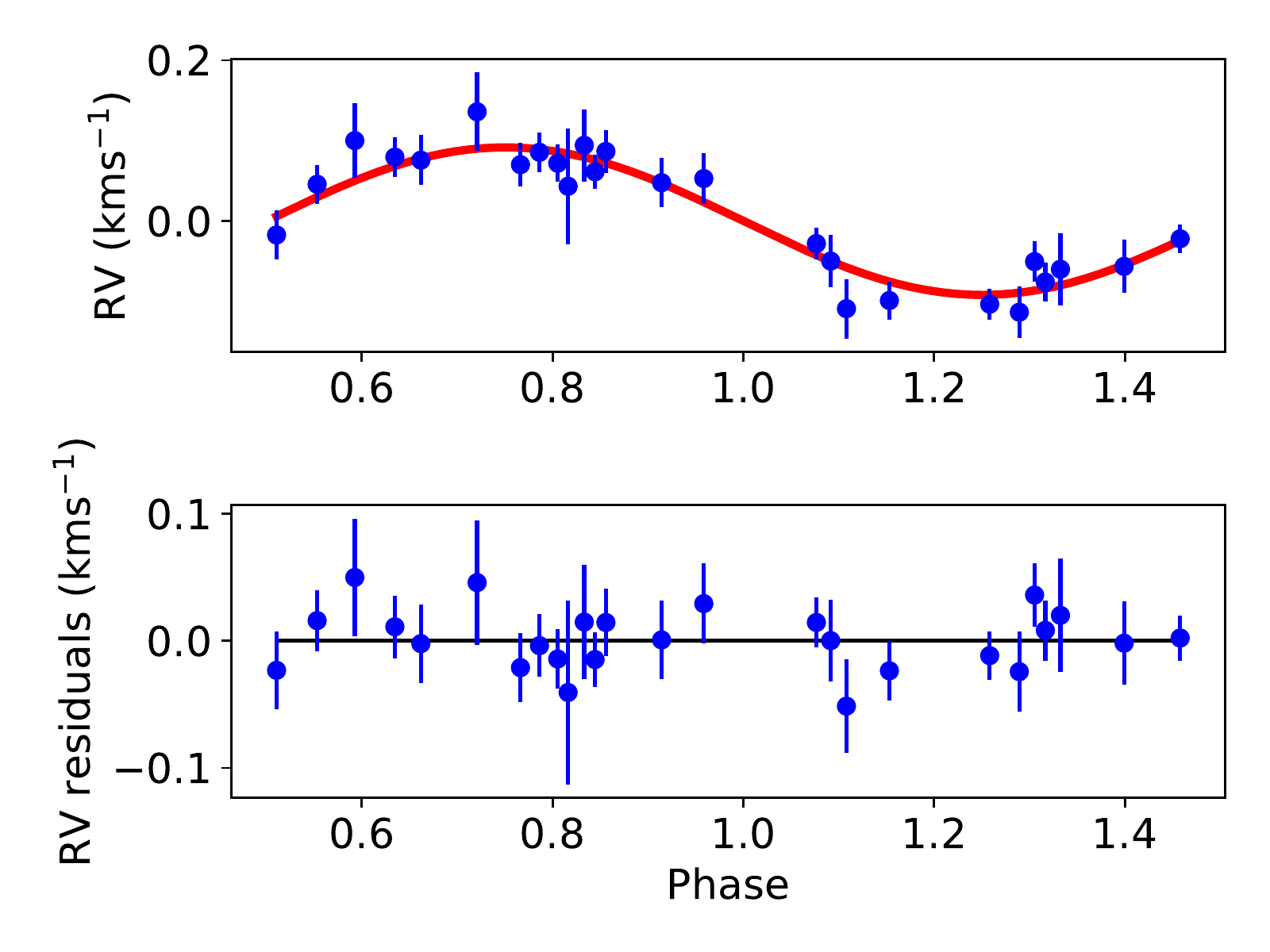}
    \caption{\textit{Upper panel}: Phase folded CORALIE RV observations of WASP-176. Data are shown in blue with the best fit model derived from MCMC analysis shown in red.
    \newline
    \textit{Lower panel}: Residuals from RV fit.}
    \label{fig:176_rv}
\end{figure}

The reduced CORALIE data are displayed in Figure \ref{fig:176_rv} along with the best fit model and residuals from our MCMC analysis (see section \ref{sec:MCMC analysis}). Data files can be found at \href{https://github.com/BenCooke95/W150-W176}{https://github.com/BenCooke95/W150-W176}.

\begin{figure}[htp]
	\includegraphics[width=\columnwidth]{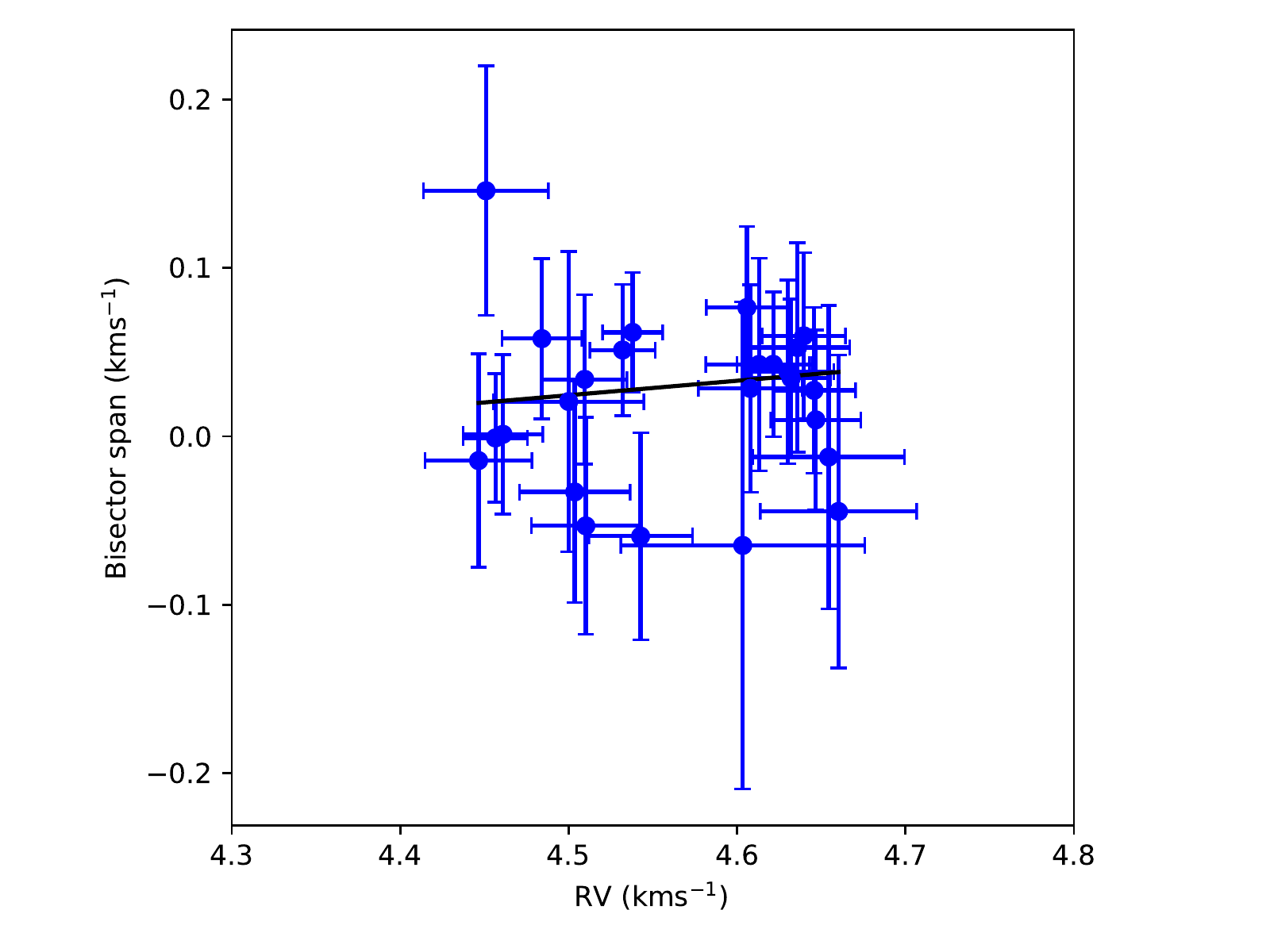}
    \caption{CORALIE RV bisector span as a function of radial velocity for WASP-176. The solid black line shows the best weighted linear fit to the data. The lack of any significant gradient supports the assumption that the RV signal is produced by a planetary companion. The aspect ratio of the two axes is unity.}
    \label{fig:176_bis_span}
\end{figure}

Additionally, the line bisector was analysed to ensure the observed signal was indeed from an orbiting body and not a blended binary \citep{Queloz2001}. Figure \ref{fig:176_bis_span} shows the results of this analysis using the bisector velocity span as a function of RV. No significant correlation is seen (evidenced by the best fit linear slope). Thus, this analysis supports the detection of a planetary companion to WASP-176. Using the Grubbs test for outliers \citep{grubbs1950} we found exactly one outlier at 95\%. This point was removed before calculating the line bisector correlation. 

\section{Photometric follow-up}
\label{sec:Photometric follow-up}

\subsection{WASP-150b}

\begin{figure}[htp]
	\includegraphics[width=\columnwidth]{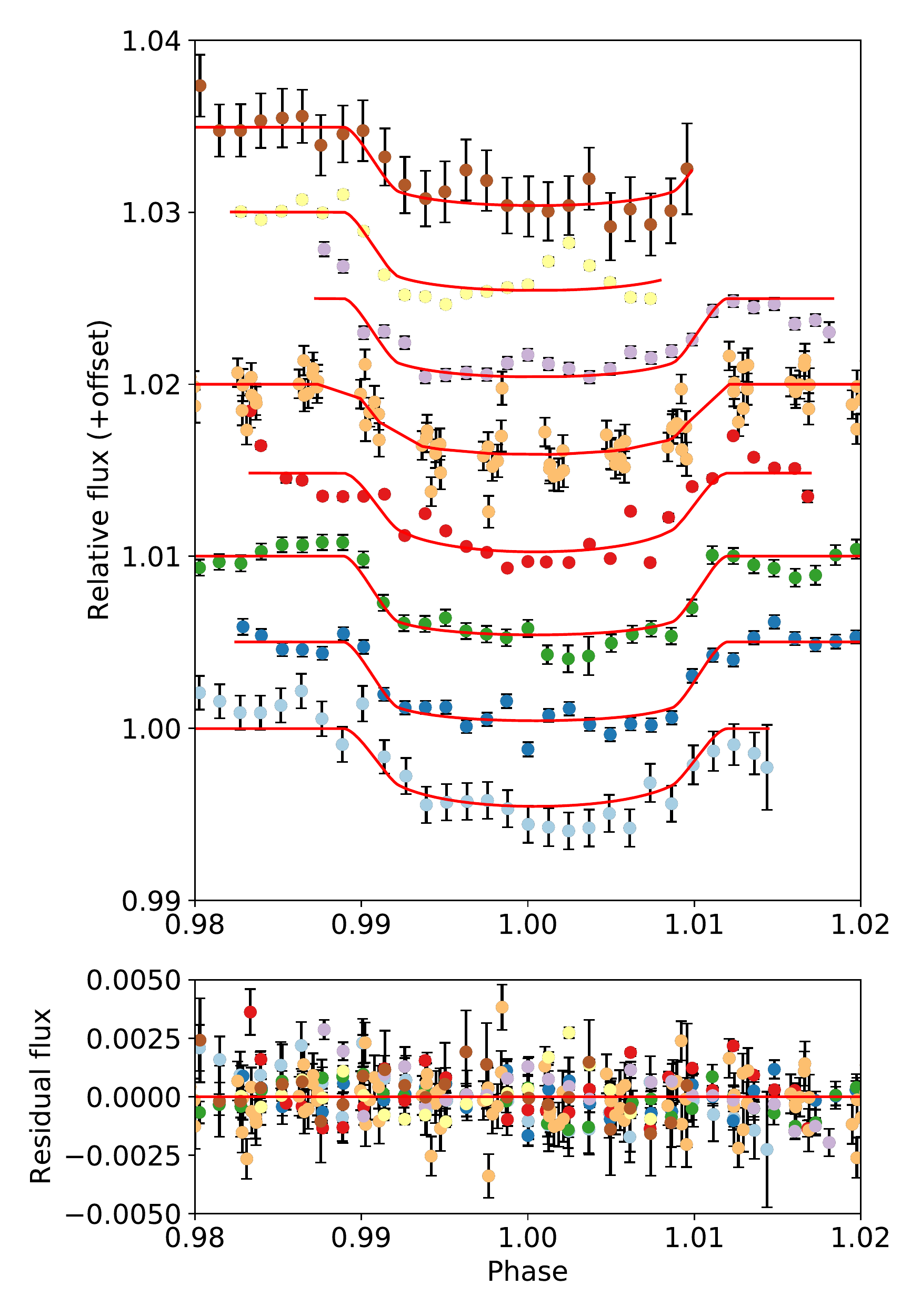}
    \caption{\textit{Upper panel}: Individual WASP-150 lightcurves binned to 10\,minutes (TESS data are left unbinned). From top down the light curves are from NITES, RISE, RISE, TESS, TCS, Cassini, CAHA and IAC80. The red curves show the best fit from MCMC.
    \newline
    \textit{Lower panel}: Best fit residuals coloured as in the upper panel.}
    \label{fig:150_individuals}
\end{figure}

\begin{figure}[htp]
	\includegraphics[width=\columnwidth]{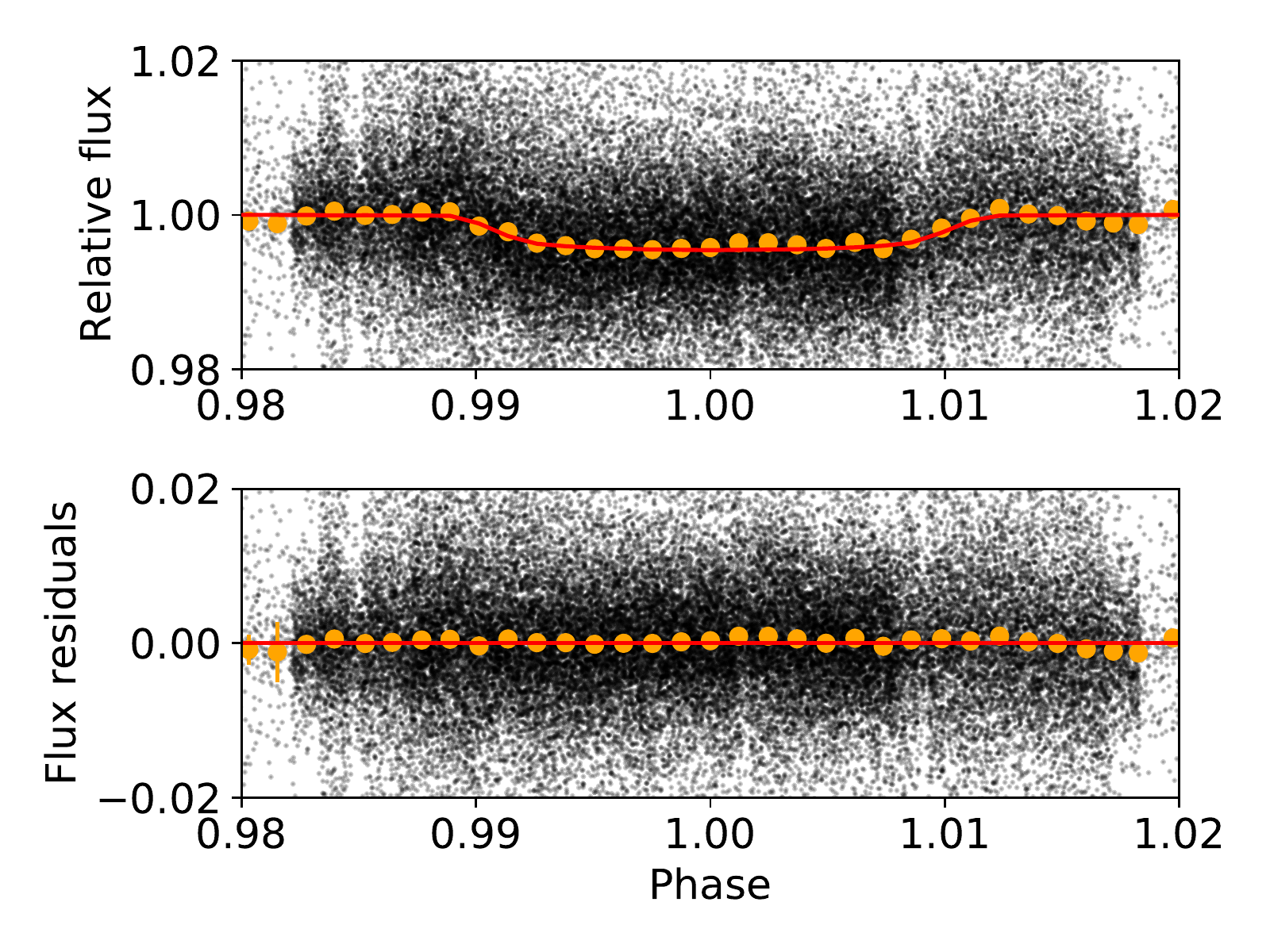}
    \caption{\textit{Upper panel}: Combined photometry data for WASP-150 binned to 10\,minutes and including best fit from MCMC.
    \newline
    \textit{Lower panel}: MCMC fit residuals.}
    \label{fig:150_all_phot}
\end{figure}

A number of telescopes were used to gather the necessary follow-up photometry for WASP-150b. These are summarised in Table \ref{tab:150Photometric follow-up} along with observation dates, photometric filters and transit notes. The following sections detail the follow-up. The lightcurves are shown in Figures \ref{fig:150_individuals} and \ref{fig:150_all_phot}. Data files can be found at \href{https://github.com/BenCooke95/W150-W176}{https://github.com/BenCooke95/W150-W176}.

\begin{table*}
	\centering
	\caption{Photometric follow-up of WASP-150}
	\label{tab:150Photometric follow-up}
	\begin{tabular}{cccccc}
		\hline
		\hline
		Instrument & Date (night of) & Filter & No. points & Average cadence (s) & Comment\\
		\hline
		NITES & 11/09/2014 & None & 599 & 26 & Noisy egress\\
		RISE & 23/05/2015 & $V + R$ & 14640 & 1 & No pre-transit OOT\\
		IAC80 & 07/07/2015 & Johnson-Bessel $B$ & 230 & 75 & Full transit observed\\
		TCS & 24/07/2015 & Johnson-Cousins $R$ & 13500 & 1 & Full transit observed\\
		RISE & 24/07/2015 & $V + R$ & 12141 & 1 & Missed egress\\
		Cassini & 27/08/2015 & Johnson $I$ & 482 & 47 & Full transit observed\\
		CAHA & 27/08/2015 & Cousins $I$ & 421 & 46 & Full transit observed\\
		\hline
	\end{tabular}
\end{table*}

\subsubsection{IAC80}
\label{sec:IAC80 150}

A full transit of WASP-150b was observed on 7 July 2015 using the CAMELOT\footnote{\url{http://vivaldi.ll.iac.es/OOCC/iac-managed-telescopes/iac80/camelot/}} (CAmara MEjorada Ligera del Observatorio del Teide) camera installed on the IAC80\footnote{\url{http://www.iac.es/OOCC/instrumentation/iac80/}} telescope at the Observatorio del Teide, Tenerife, Spain. CAMELOT contains a $2048 \times 2048$ back-illuminated e2v CCD, providing a field-of-view of $10.4^{\prime}\times10.4^{\prime}$, with a pixel scale of $0.304^{\prime\prime}$.


Data were reduced using standard routines of {\scshape iraf} \citep{tody86, tody93}. All images were bias and flat subtracted and differential photometry was carried out using IDL {\scshape daophot}-Type Photometry Procedures\footnote{\url{https://idlastro.gsfc.nasa.gov/contents.html}} \citep{stetson87}. Among the several stars appearing on the FOV of the camera, those showing less dispersion were selected to produce an average reference star and obtain the final light curve. We used a fixed aperture radius of 13 pixels, which minimised the RMS scatter in the out of transit data.

\subsubsection{CAHA 1.23~m}
\label{sec:CAHA 150}

The transit on 27 August 2015 was observed with the DLR-MKIII camera fed by the Zeiss 1.23\,m CAHA telescope. During the observations the sky was clear besides a couple of intervals when some thin clouds passed in front of the target. The observations were carried out using the defocusing technique, which allowed the use of longer exposures compared to in focus observations, without the risk of saturation \citep{Southworth2009}.

The data reduction was performed using standard methods making use of the {\scshape defot} pipeline \citep{Southworth2009,Southworth2014}. In brief, each scientific image was calibrated using a master-bias and a master-flat. The fluxes detected from the target and comparison stars were then obtained via aperture photometry, selecting the aperture sizes that minimised the scatter of the light curves. A relative-flux light curve was then obtained for the target star relative to an optimally-weighted composite comparison star constructed from the light curves of comparison stars present in the same field of view, to account for atmospheric and instrumental changes occurring during the observations.

\subsubsection{Cassini 1.52~m}
\label{sec:Cassini 150}


On 27 August 2015 WASP-150b was simultaneously observed with the Cassini 1.52\,m telescope at the Astronomical Observatory of Bologna in Loiano (Italy), thus performing the two-site observational strategy \citep{Ciceri2013}. The Cassini has a focal ratio of f/8, a focal length of 12\,m, and is equipped with a back-illuminated CCD with $1300\times1340$ pixels and a pixel size of 20\,$\mu$m. A focal reducer makes the telescope an f/5, so that its plate scale is $0.58^{\prime\prime}$ pixel$^{-1}$ and the FOV is $13^{\prime}\times12.6^{\prime}$.
The autoguided observations 
were performed with the defocusing technique 
(to improve the photometric precision), 
though the level of defocusing was limited in order to avoid blending from a 
fainter star 
a few arcsec away.
The data were reduced using the {\scshape defot} code, as described in the preceding section.

\subsubsection{RISE}
\label{sec:RISE 150}

Two partial transits of WASP-150b were observed on 23 May 2015 and 24 July 2015 with the RISE\footnote{\url{https://telescope.livjm.ac.uk/TelInst/Inst/RISE/}} (Rapid Imaging Search for Exoplanets) optical camera, installed on the 2-meter robotic Liverpool Telescope \citep{steele08, gibson08} at the Observatorio del Roque de los Muchachos, La Palma, Spain. The camera is a frame transfer e2v CCD of $1024 \times 1024$ pixels, which has a FOV of $9.2^{\prime} \times 9.2^{\prime}$.


Bias and flat reduced images were provided by the telescope pipeline. The light curves were extracted following the standard procedures described in section \ref{sec:IAC80 150}, using a fixed aperture radii of 5.5 and 7 pixels, for the first and second night respectively.

\subsubsection{TCS}
\label{sec:TCS 150}

On 24 July 2015, WASP-150b was simultaneously observed with the Telescopio Carlos S\'anchez\footnote{\url{http://www.iac.es/OOCC/instrumentation/telescopio-carlos-sanchez/}} (TCS) a 1.52-metre telescope installed at the Observatorio del Teide, once again performing the two-site observational strategy \citep{ciceri13}. We used the \textsc{Wide FastCam} camera, a $1024 \times 1024$ pixels EMCCD detector, coupled to an optical design \citep{murga14} that is able to provide $\sim 8^{\prime} \times 8^{\prime}$ FOV. This camera provides small readout times and low electronic noise, which allow us to precisely measure exoplanet transit timings.


All images were bias and flat subtracted, and light curves were extracted using similar procedures to those described in section \ref{sec:IAC80 150}. In this case, a fixed radius aperture of 11.5 pixels was selected.

\subsubsection{NITES}
\label{sec:NITES 150}

A transit of WASP-150b was obtained on 11 September 2014 using the Near Infra-red Transiting ExoplanetS telescope \citep[NITES,][]{McCormac2014} on La Palma.
The data were reduced in {\scshape Python} using {\scshape ccdproc} \citep{Craig2015}. A master bias, dark and flat was created using the standard process. Non-variable nearby comparison stars were selected by hand, and aperture photometry extracted using {\scshape sep} \citep{Barbary2016,Bertin1996}.

\subsubsection{TESS}
\label{sec:TESS}

The Transiting Exoplanet Survey Satellite \citep[TESS,][]{2015JATIS...1a4003R} observed WASP-150b during its northern hemisphere campaign. WASP-150b was observed in the full frame images of sectors 14 and 20. To account for the blending in the TESS lightcurve (TESS pixels are 21$^{\prime\prime}$ square) this data was included with a variable dilution term in the MCMC modelling for WASP-150b. This light curve was extracted from the full frame images using Eleanor extraction pipeline \citep{2019PASP..131i4502F} utilising the background subtraction and systematics removal packages.

\subsection{WASP-176b}

\begin{figure}[htp]
	\includegraphics[width=\columnwidth]{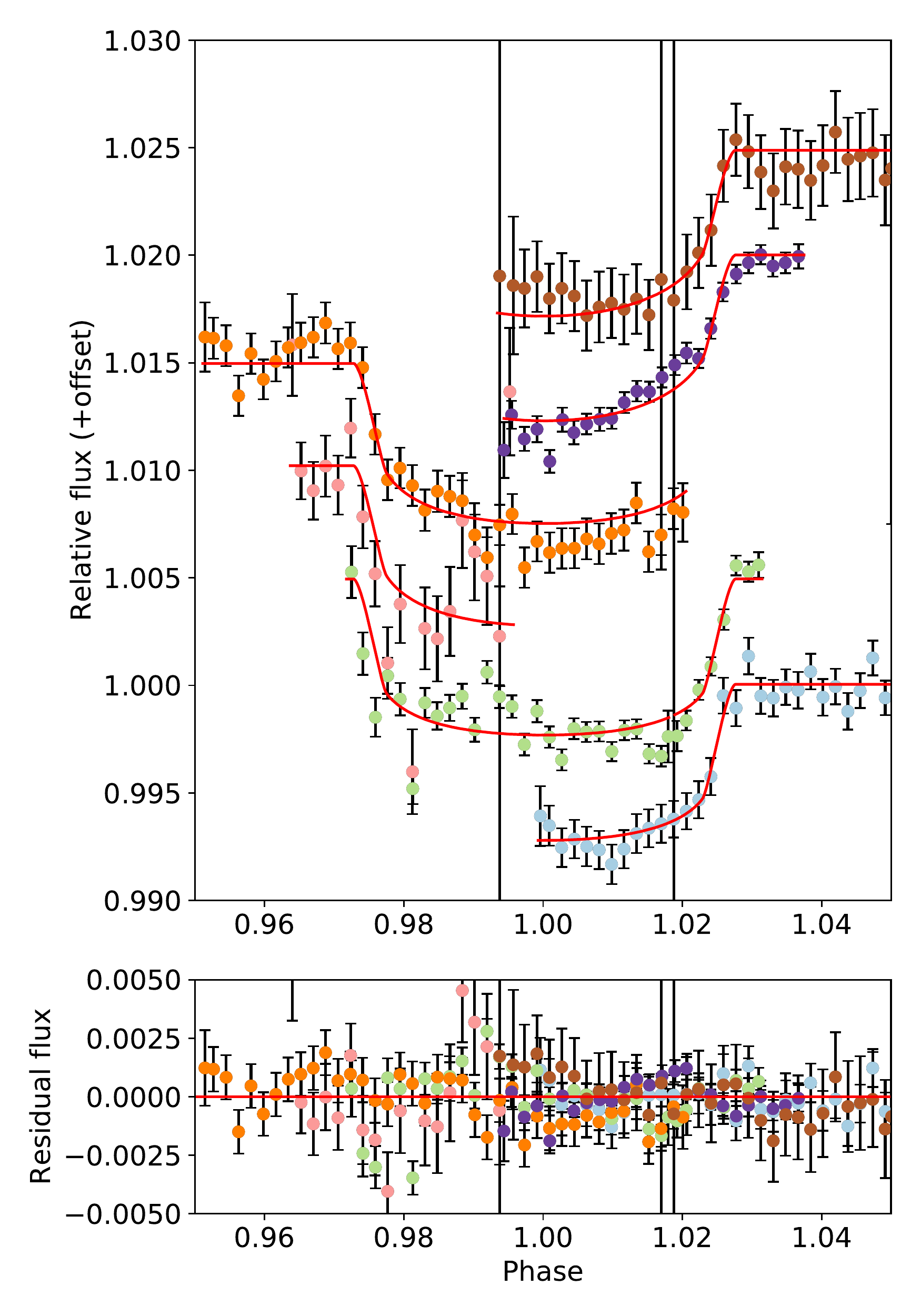}
    \caption{\textit{Upper panel}: Individual WASP-176 lightcurves binned to 10\,minutes. From top down the light curves are from NITES, TRAPPIST, NITES, SPECULOOS, Cassini and RISE2. The red curves show the best fit from MCMC.
    \newline
    \textit{Lower panel}: Best fit residuals coloured as in the upper panel.}
    \label{fig:176_individuals}
\end{figure}

\begin{figure}[htp]
	\includegraphics[width=\columnwidth]{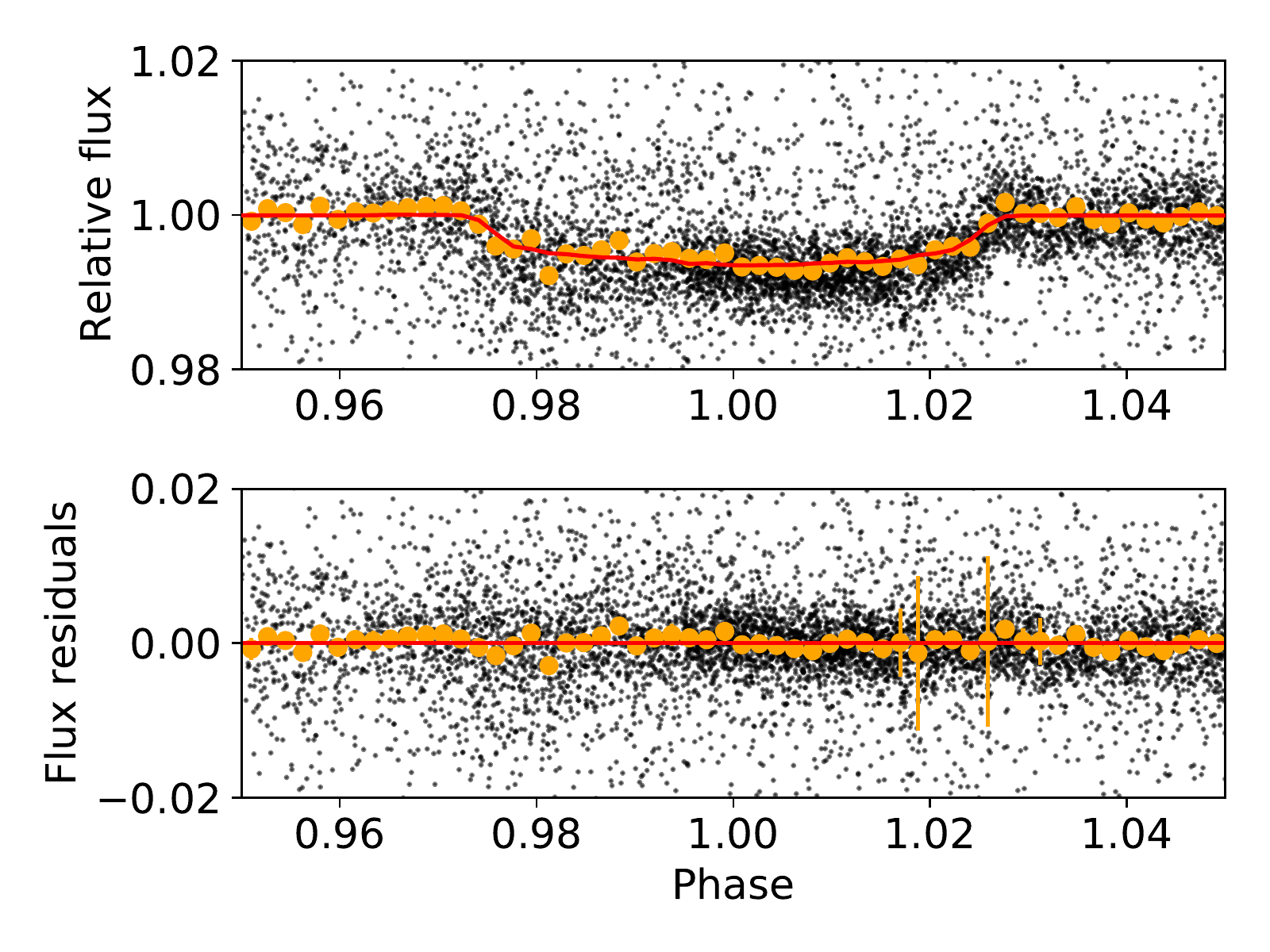}
    \caption{\textit{Upper panel}: Combined photometry data for WASP-176 binned to 10\,minutes and including best fit from MCMC.
    \newline
    \textit{Lower panel}: MCMC fit residuals.}
    \label{fig:176_all_phot}
\end{figure}

To gather the necessary follow-up photometry for WASP-176b, a number of telescopes were used. These are summarised in Table \ref{tab:176Photometric follow-up} along with observation dates, photometric filters and transit notes. The following sections detail the follow-up. The lightcurves are shown in Figures \ref{fig:176_individuals} and \ref{fig:176_all_phot}. Data files can be found at \href{https://github.com/BenCooke95/W150-W176}{https://github.com/BenCooke95/W150-W176}.

\begin{table*}
	\centering
	\caption{Photometric follow-up of WASP-176}
	\label{tab:176Photometric follow-up}
	\begin{tabular}{cccccc}
		\hline
		\hline
		Instrument & Date (night of) & Filter & No. points & Average cadence (s) & Comment\\
		\hline
		SPECULOOS-Io & 14/06/2018 & Sloan $z^{\prime}$ & 1127 & 21 & Missed egress\\
		TRAPPIST-North & 26/06/2018 & $I+z$ & 708 & 28 & Full but no out of transit (meridian flip)\\
        Cassini & 30/06/2018 & Johnson $V$ & 123 & 119 & Missed ingress (cloud)\\
        NITES & 04/08/2018 & Johnson-Bessel $R$ & 340 & 32 & Missed ingress\\
        NITES & 08/08/2018 & Johnson-Bessel $I$ & 676 & 32 & Missed egress\\
        RISE-2 & 24/09/2018 & $V+R$ & 1929 & 11 & Missed ingress\\
		\hline
	\end{tabular}
\end{table*}

\subsubsection{Cassini 1.52~m}
\label{sec:Cassini 176}

A partial transit of WASP-176b was recorded with the Cassini 1.52\,m telescope on 30 June 2018. Unfortunately, cloud prevented observing the start of the transit. 
The observations were performed by defocusing the telescope for improving the photometric precision and using autoguiding. The data were reduced as discussed in Section \ref{sec:Cassini 150}. We detrended the light curve to remove slow instrumental and astrophysical trends, by fitting a straight line to the out-of-transit data.

\subsubsection{NITES}
\label{sec:NITES176}

A total of 2 transits were obtained using NITES \citep{McCormac2014} on La Palma. 
The data were reduced in {\scshape Python} using {\scshape ccdproc} \citep{Craig2015}. A master bias, dark and flat were created using the standard process on each night. Twenty one images of each type were used for the master calibration frames. Non-variable nearby comparison stars were selected by hand, and aperture photometry extracted using {\scshape sep} \citep{Barbary2016,Bertin1996}. The aperture photometry radii were chosen to minimise the dispersion in the data points out of transit.

\subsubsection{RISE-2}
\label{sec:RISE-2 176}

WASP-176 was observed 
with RISE-2 mounted on the 2.3\,m telescope situated at Helmos observatory in Greece. RISE-2 has a CCD size of 1K$\times$1K with a pixel scale of $0.51^{\prime\prime}$ and a field of view of $9^{\prime}\times9^{\prime}$ \citep{Boumis2010}.
The data were reduced using master bias and flat frames, created using the standard process on each night. Non-variable nearby comparison stars were selected by hand and aperture photometry extracted using {\scshape sep} \citep{Barbary2016,Bertin1996}. The aperture photometry radii were chosen to minimise the dispersion in the data points out of transit.


\subsubsection{TRAPPIST}
\label{sec:TRAPPIST 176}

TRAPPIST-North \citep{Gillon2017,Barkaoui2018} observed one full transit of WASP-176b on 26 June 2018. 
TRAPPIST-North is a 60\,cm robotic telescope installed in spring 2016 at Oukaimeden Observatory in Morocco. 
TRAPPIST-North is a northern twin of TRAPPIST-South \citep{Jehin2011,Gillon2011}. TRAPPIST-North has an f/8 Ritchey-Chretien optical design. It is equipped with a thermoelectrically-cooled $2048\times2048$ deep-depletion Andor iKon-L CCD camera that has a pixel size of 13.5\,$\mu$m, which translates into a $0.60^{\prime\prime}$ pixel$^{-1}$ image scale and a field of view of $19.8^{\prime}\times19.8^{\prime}$. 
Data reduction consisted of standard calibration steps (bias, dark and flat-field corrections) and subsequent aperture photometry using {\scshape iraf/daophot} \citep{Tody1986}. Extraction of fluxes of selected stars using aperture photometry was performed with {\scshape iraf/daophot} (as described in \citealt{Gillon2013}).

During the TRAPPIST-North observations of WASP-176 the telescope underwent a meridian flip at JD 2458296.6355. To counter this problem the pre- and post-meridian flip data are treated separately.

\subsubsection{SPECULOOS}
\label{sec:SPECULOOS 176}

One partial transit of WASP-176b was observed with SPECULOOS-Io, one of the four telescopes of the SPECULOOS-South facility \citep{Gillon2018,Delrez2018,Burdanov2018,Jehin2018}, which is located at ESO Paranal Observatory (Chile). Each telescope is a robotic Ritchey-Chretien (f/8) telescope of 1m diameter. They are equipped with Andor iKon-L Peltier-cooled deeply depleted 2K$\times$2K CCD cameras, with good sensitivities in the very-near-infrared up to 1\,$\mu$m. The field of view of each telescope is $12^{\prime}\times12^{\prime}$ and the pixel scale is $0.35^{\prime\prime}$ pixel$^{-1}$. 
The calibration and photometric reduction of the data were performed as described in \citet{Gillon2013}.

\section{High-spatial-resolution follow-up}
\label{sec:imaging}

WASP-150 was observed on four occasions: 15 November 2015, 9 March 2016, 10 March 2016 and 6 May 2016; using the \textsc{FastCam} camera \citep{oscoz08} installed on the 1.52\,m Telescopio Carlos S\'anchez (TCS) located at the Observatorio del Teide, Tenerife, Spain. \textsc{FastCam} is a EMCCD camera with $512 \times 512$ pixels, with a physical pixel size of 16\,$\mu$m, which makes a FOV of $21.2^{\prime\prime} \times 21.2^{\prime\prime}$. Thanks to the very low noise and fast readout speed of the EMCCD array, this camera is appropriate for Lucky Imaging (LI) observations.

During each of the four observing nights, 10\,000 individual frames of WASP-150 were collected in the Johnson-Cousins $I$ filter, except on the 15 November 2015 night, for which clear filter was used due to the dusty weather conditions (\textit{calima}). Each individual frame had an exposure time of 50\,ms. In total, 500 seconds and 1\,500 seconds of data were gathered of WASP-150, with the clear and $I$-band filters, respectively.

\begin{figure}[htp]
\includegraphics[width=\columnwidth]{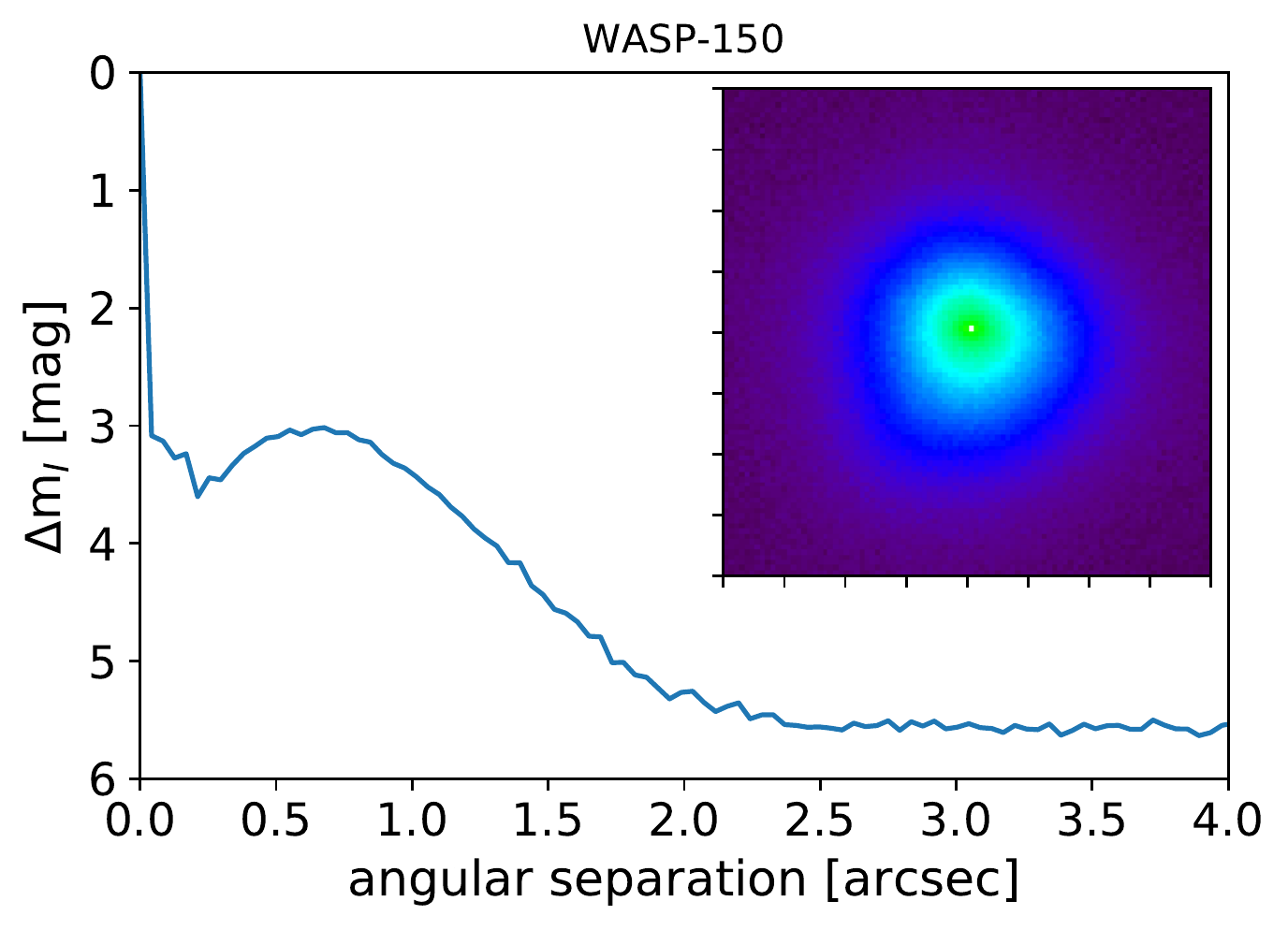}
\caption{I-band magnitude contrast as a function of angular separation up to $4.0^{\prime\prime}$ of WASP-150. The solid line indicate the $5$-$\sigma$ detection limit for primary star. The inset shows the $2.0 \times 2.0^{\prime\prime}$ combined image of WASP-150. North is up and East is left.}
\label{lucky}
\end{figure}

Using the \textsc{FastCam} dedicated software developed at the Universidad Polit\'ecnica de Cartagena \citep{labadie10, jodar13}, each individual frame was bias-subtracted, aligned and co-added and then processed in order to construct a high resolution, long-exposure image. For each night's data set, we took a high resolution image constructed by co-addition of the best $30\%$ of images, thus making a 150 seconds total exposure time. No close companion was detected, only a $\Delta m_I = 2.17 \pm 0.03$ mag fainter star at a distance of $10.58 \pm 0.05$ arcsec. Figure \ref{lucky} shows the contrast light curve that was computed based on the scatter within the annulus as a function of angular separation from the target centroid \citep[see, e.g.,][]{gauza15}.

\section{Results}
\label{sec:Results}

\subsection{WASP-150b}

\subsubsection{Stellar parameters}

To determine the stellar parameters of WASP-150 a spectral analysis of the SOPHIE spectra was carried out. The results of this analysis are shown in Table \ref{tab:Stellar parameters}.

\begin{table}
	\centering
	\caption{Stellar parameters from spectral analysis}
	\label{tab:Stellar parameters}
    \begin{threeparttable}
	\begin{tabular}{ccc}
		\hline
		\hline
		Parameter (unit) & WASP-150 & WASP-176\\
		\hline
		Spectral type & \starclassspecA & \starclassspecB\\
		$T_{eff}$ (K) & \starteffspecA & \starteffspecB\\
		$\log{g}$ & \starloggspecA & \starloggspecB\\
        $[Fe/H]$ & \starfehspecA & \starfehspecB\\
        $v\sin{i}$ (kms$^{-1}$) & \starvsinispecA\tnote{a} & \starvsinispecB\tnote{b}\\
		\hline
	\end{tabular}
    \begin{tablenotes}
            \item[a] Assuming a microturbulance of $0.78\pm0.05$ kms$^{-1}$ from \citet{Doyle2013} calibration.
            \item[b] Assuming a macroturbulance of $5.1\pm0.7$ kms$^{-1}$ from \citet{Doyle2013} calibration.
        \end{tablenotes}
\end{threeparttable}
\end{table}

Additionally, the second data release of Gaia \citep{Gaia2016,Gaia2018} provided some stellar parameters presented in Table \ref{tab:Gaia}. Gaia was also searched for close companions of WASP-150. No significant companion was found.

\begin{table}
	\centering
	\caption{Stellar parameters from Gaia DR2}
	\label{tab:Gaia}
	\begin{tabular}{ccc}
		\hline
		\hline
		Parameter (unit) & WASP-150 & WASP-176\\
		\hline
		Parallax (mas) & \starparallaxgaiaA & \starparallaxgaiaB\\
        Distance (pc) & \stardistgaiaA & \stardistgaiaB\\
        PM R.A. (mas yr$^{-1}$) & \starpmragaiaA & \starpmragaiaB\\
        PM Dec. (mas yr$^{-1}$) & \starpmdecgaiaA & \starpmdecgaiaB\\
        $T_{eff}$ (K) & \starteffgaiaA & \starteffgaiaB\\
        $R_\star$ ($\rm R_{\odot}$) & \starradiusgaiaA & \starradiusgaiaB\\
        $L_\star$ ($\rm L_{\odot}$) & \starlumgaiaA & \starlumgaiaB\\
		\hline
	\end{tabular}
\end{table}

The open-source stellar modelling code {\scshape bagemass}\footnote{\url{https://sourceforge.net/projects/bagemass/}} \citep{Maxted2015a} was then used to estimate the age and mass of WASP-150. {\scshape bagemass} uses the {\scshape garstec} stellar evolution code \citep{Weiss2008} to calculate model grids of individual stars. A Bayesian method then samples the posterior distributions on mass and age. The results of this analysis are presented in Table \ref{tab:BAGEMASS properties}.


\begin{table}
	\centering
	\caption{Stellar properties from {\scshape bagemass}}
	\label{tab:BAGEMASS properties}
	\begin{tabular}{ccc}
		\hline
		\hline
		Parameter (unit) & WASP-150 & WASP-176\\
		\hline
		Mass ($\rm M_\odot$) & \starmassBAGEMASSA & \starmassBAGEMASSB\\
		$\tau_{iso}$ (Gyr) & \starageisoA & \starageisoB\\
		$[Fe/H]_{init}$ & \starfehBAGEMASSA & \starfehBAGEMASSB\\
		\hline
	\end{tabular}
\end{table}

\begin{figure}[htp]
	\includegraphics[width=\columnwidth]{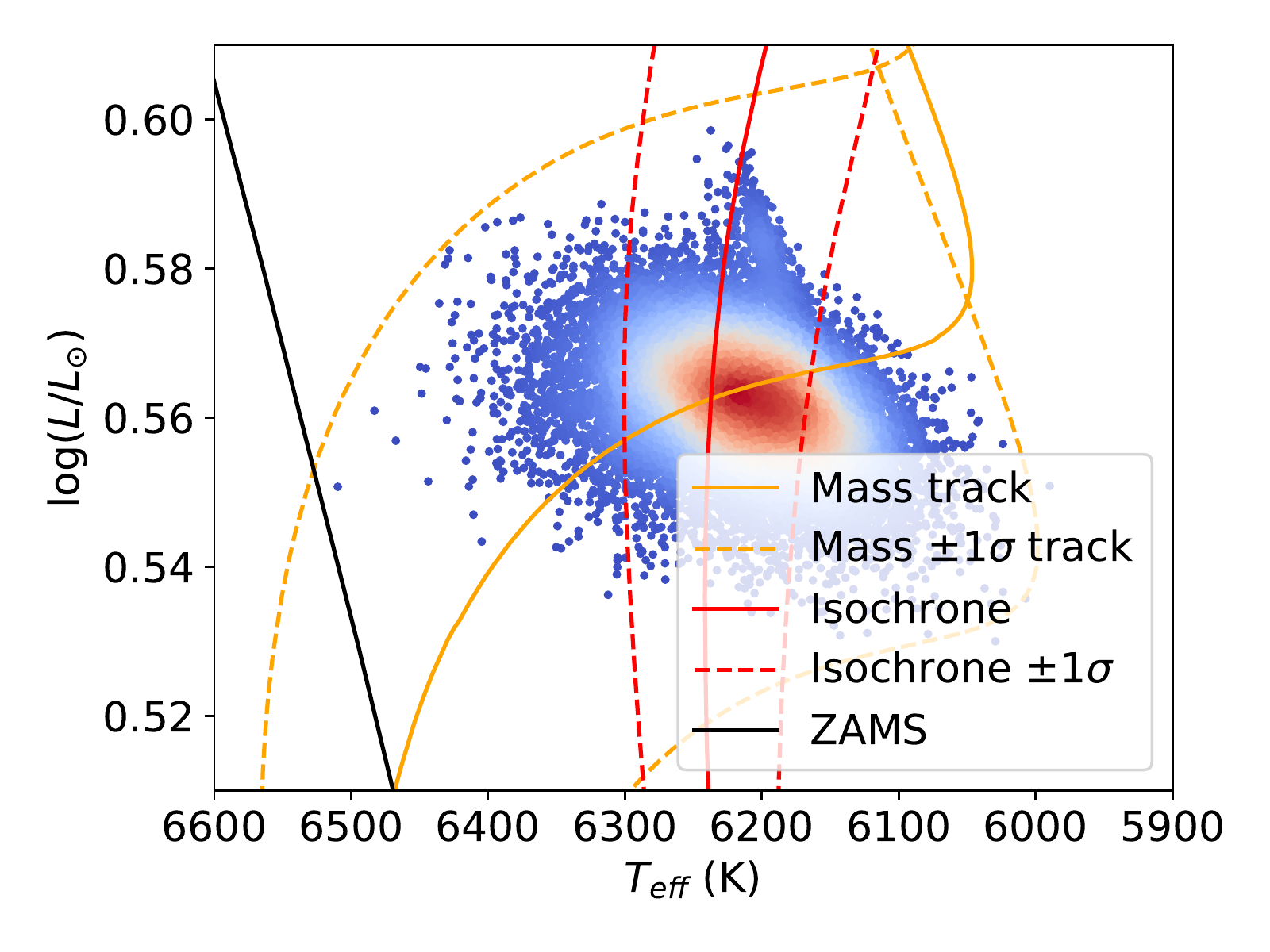}
    \caption{{\scshape bagemass} analysis of WASP-150. The black line shows the Zero Age Main Sequence (ZAMS). The orange line shows the best fit mass track with the dotted lines showing the $\pm1\sigma$ mass tracks. The red line shows the isochrone relating to the best fit age with the dotted lines showing the $\pm1\sigma$ isochrones. Finally the density of EXOFASTv2 samples is shown in the colour scale of the plotted posterior distribution.}
    \label{fig:150_hr}
\end{figure}

The best-fit evolutionary track and isochrone produced from the {\scshape bagemass} analysis are shown in Figure \ref{fig:150_hr} along with the $1\sigma$ uncertainties. This plot also includes the posterior distribution produced by the EXOFASTv2 analysis (see section \ref{sec:EXOFASTv2 analysis}).

Figure \ref{fig:150_hr} includes a small collection of points above the main area of convergence. These data are not affected by increasing the run-time of the MCMC analysis or by increasing the burn-in period thus we do not believe them to be an artefact of unfinished fitting. However, since the significance of this region is $\lesssim$\,10 times lower than the peak we do not find it impactful.

\subsubsection{EXOFASTv2 analysis}
\label{sec:EXOFASTv2 analysis}

To perform simultaneous fitting of the SuperWASP detection, the SOPHIE RVs and the follow-up photometry we used the fitting code EXOFASTv2 \citep{Eastman2017,Eastman2019}. This tool is designed to fit all the available data and ensure consistency between derived stellar and planetary parameters. EXOFASTv2 explores the given parameter space through a differential evolution Markov chain method using 30\,000 steps. We use the Gelman-Rubin statistic \citep{Gelman2003} to check the mixing of the chains as proposed by \citet{Ford2006}. We fit a total of 50 free parameters, these are the parameters presented in Table \ref{tab:MCMC parameters} as well as limb darkening parameters for each band used and offset baselines and variances to account for any errors in normalisation.

For the EXOFASTv2 input parameters we take period and epoch from the initial SuperWASP discovery photometry. Metallicity and effective temperature are taken from the results of spectroscopic analysis. Additionally, we use stellar radius and luminosity as well as parallax and distance from Gaia DR2. Finally we impose a prior on the V-band extinction from \citet{Schlegel1998} and \citet{Schlafly2011} which are used to model the stellar properties through SED fitting. Within EXOFASTv2 we use the MESA Isochrones and Stellar Tracks \citep[MIST,][]{Dotter2016,Choi2016} to model the star. This produces an age of $2.18^{+0.58}_{-0.65}$\,Gyr, which is in reasonable agreement with the {\scshape bagemass} 
age in Table \ref{tab:BAGEMASS properties}. The best fit values, along with uncertainties, are presented in Table \ref{tab:MCMC parameters}.

\subsection{WASP-176b}

\subsubsection{Stellar parameters}
\label{sec:Stellar parameters 176}

Similar to WASP-150, the stellar parameters of WASP-176 are calculated via a spectral analysis of the CORALIE spectra. The results of this analysis are shown in Table \ref{tab:Stellar parameters}.

Once again, the second data release of Gaia \citep{Gaia2016,Gaia2018} provided some additional stellar parameters presented in Table \ref{tab:Gaia}. Gaia was also searched for close companions of WASP-176.

To determine the age of WASP-176 we again ran {\scshape bagemass} and show the results in Table \ref{tab:BAGEMASS properties}.

\begin{figure}[htp]
	\includegraphics[width=\columnwidth]{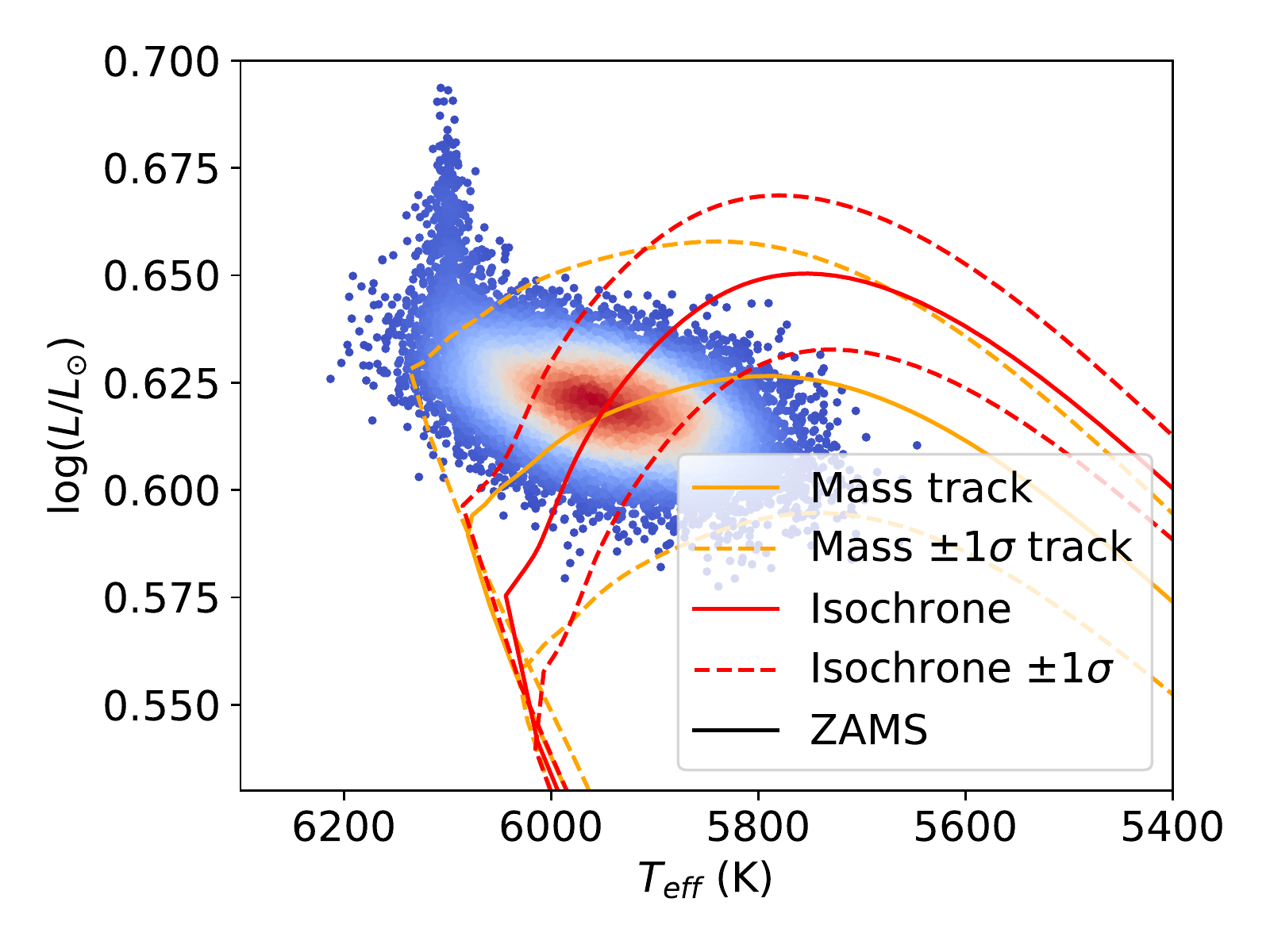}
    \caption{{\scshape bagemass} analysis of WASP-176. The black line shows the Zero Age Main Sequence (ZAMS). The orange line shows the best fit mass track with the dotted lines showing the $\pm1\sigma$ mass tracks. The red line shows the isochrone relating to the best fit age with the dotted lines showing the $\pm1\sigma$ isochrones. Finally the density of MCMC samples is shown in the colour scale of the plotted posterior distribution.}
    \label{fig:176_hr}
\end{figure}

The best-fit evolutionary track and isochrone produced from the {\scshape bagemass} analysis are shown in Figure \ref{fig:176_hr} along with the $1\sigma$ uncertainties. This plot also includes the posterior distribution produced by the MCMC analysis (see section \ref{sec:MCMC analysis}).

Once again we see that Figure \ref{fig:176_hr} includes a few points above the main peak. As before we find that these data are not affected by increasing the run-time of the MCMC analysis or by increasing the burn-in period so, again, we do not believe them to be an artefact of unconverged fitting. Since the significance of this region is $\sim$10 times lower than the peak we do not find it impactful.

\subsubsection{EXOFASTv2 analysis}
\label{sec:MCMC analysis}

As for WASP-150b we once again use EXOFASTv2 to derive the quoted parameters for WASP-176b. We use the same number of steps, 30\,000, and utilise the Gelman-Rubin statistic to check the mixing of the chains. We fit 43 free parameters, these are the parameters presented in Table \ref{tab:MCMC parameters} as well as limb darkening parameters for each band used and offset baselines and variances to account for any errors in normalisation. In the same way as for WASP-150b we take the EXOFASTv2 input parameters from a combination of the WASP discovery photometry (period and epoch), spectroscopic analysis (metallicity and effective temperature) and Gaia DR2 (stellar radius, luminosity, parallax and distance). Additionally we impose a V-band extinction prior from \citet{Schlegel1998} and \citet{Schlafly2011}. We use the MESA Isochrones and Stellar Tracks \citep[MIST,][]{Dotter2016,Choi2016} to model the star. This produces an age of $3.69^{+1.9}_{-0.83}$\,Gyr, which is in good agreement with the {\scshape bagemass} 
age in Table \ref{tab:BAGEMASS properties}. The best fit values, along with uncertainties, are presented in Table \ref{tab:MCMC parameters}.

The WASP-176b analysis was first conducted allowing for an eccentric orbit. This resulted in $\chi^2_{ecc} = 11.423$ Repeating the analysis, this time forcing a circular orbit gives $\chi^2_{circ} = 11.358$. Since the discrepancy between these values is negligible it was decided there was insufficient evidence to support eccentricity and thus a circular orbit was assumed.

\renewcommand{\arraystretch}{1.75}
\begin{table}
	\centering
	\caption{System parameters from MCMC analysis}
	\label{tab:MCMC parameters}
	\begin{tabular}{ccc}
		\hline
		\hline
		Parameter (unit) & WASP-150b & WASP-176b\\
		\hline
		$T_0$ (HJD) & \epochA & \epochB\\
        $P$ (days) & \periodA & \periodB\\
        $\Delta F$ & \deltafluxA & \deltafluxB\\
        $T_{14}$ (days) & \transitdurationA & \transitdurationB\\
        $b$ & \impactA & \impactB\\
        $i$ ($^{\circ}$) & \inclinationA & \inclinationB\\
        $e$ & \eccentricityA & \eccentricityB\\
        $M_\star$ ($\rm M_{\odot}$) & \starmassA & \starmassB\\
        $R_\star$ ($\rm R_{\odot}$) & \starradiusA & \starradiusB\\
        $\rho_\star$ ($\rm \rho_{\odot}$) & \stardensityA & \stardensityB\\
        $\log{g_\star}$ (cgs) & \starloggA & \starloggB\\
        $T_{eff}$ (K) & \starteffA & \starteffB\\
        $[Fe/H]$ & \starfehA & \starfehB\\
        $M_{pl}$ ($\rm M_J$) & \planetmassA & \planetmassB\\
        $R_{pl}$ ($\rm R_J$) & \planetradiusA & \planetradiusB\\
        $\rho_{pl}$ ($\rm \rho_J$) & \planetdensityA & \planetdensityB\\
        $\log{g_{pl}}$ (cgs) & \planetloggA & \planetloggB\\
        $T_{pl}$ (K) & \planettempA & \planettempB\\
        $a$ (AU) & \planetsepA & \planetsepB\\
		\hline
	\end{tabular}
\end{table}

\section{Discussion and conclusions}
\label{sec:Discussion and conclusions}

\subsection{WASP-150b}

\begin{figure}[htp]
	\includegraphics[width=\columnwidth]{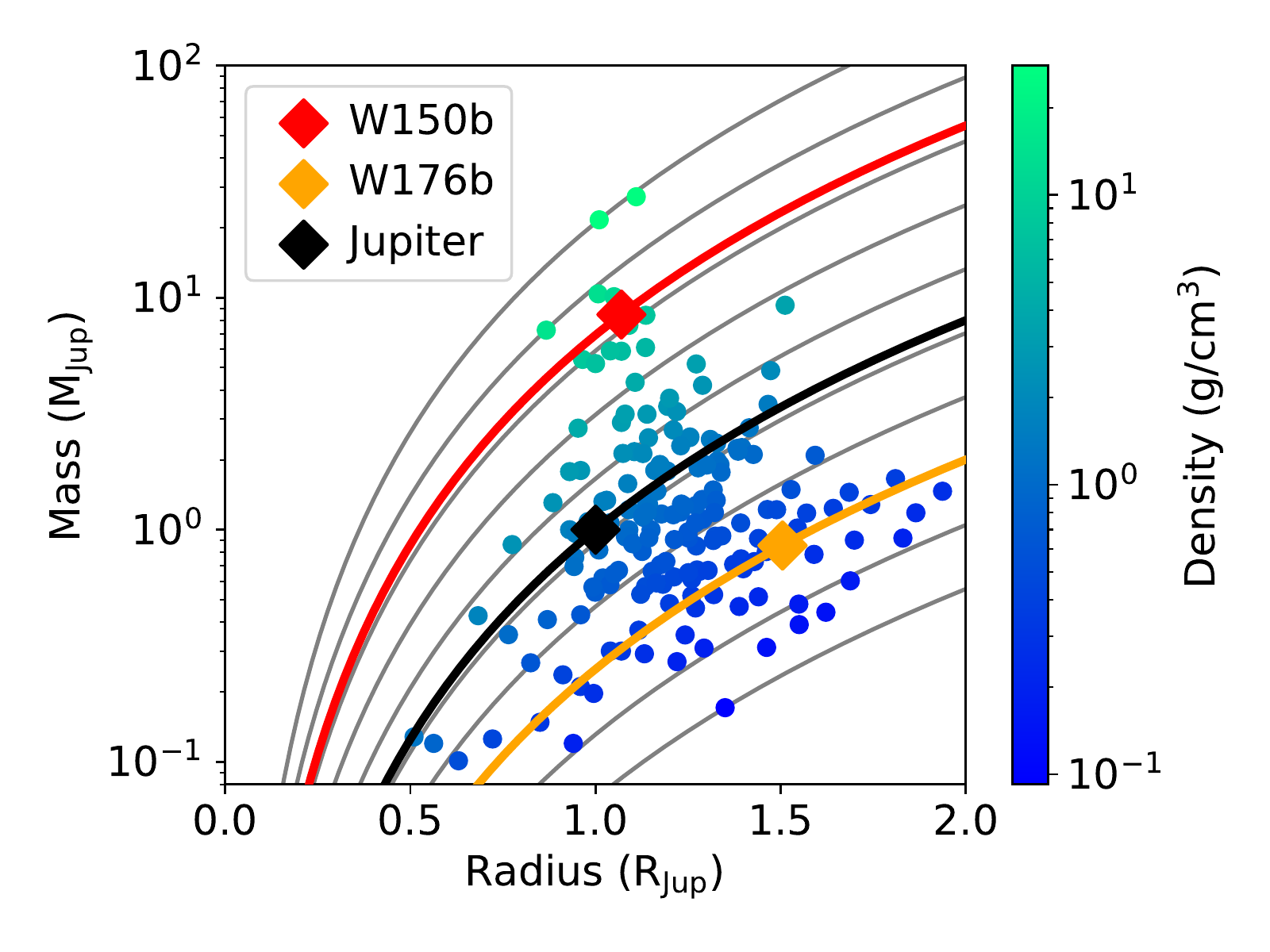}
    \caption{Scatter plot of all confirmed exoplanets from the NASA exoplanet archive with radius and mass known to an accuracy of $\leq$ 10\% and periods of $\leq$ 10 days. The points are then coloured by density in cgs units. WASP-150b is shown as a red diamond and Jupiter is shown as a black diamond for reference (WASP-176b is shown as an orange diamond). The plot also includes lines of constant density to guide the eye.}
    \label{fig:density plot}
\end{figure}

WASP-150b is a high density hot Jupiter on a \periodA\,day orbit around its \starclassspecA~host. With a mass of of \planetmassA\,$\rm M_J$ and a radius of \planetradiusA\,$\rm R_J$ WASP-150b has a density of \planetdensityA\,$\rm \rho_J$ placing it amongst the highest density planets known. Figure \ref{fig:density plot} shows a plot of all the confirmed exoplanets from the NASA exoplanet archive with radius and mass known to an accuracy of $\leq$ 10\% and periods of $\leq$ 10 days. Though large, this density is in line with expectations based on the planets mass \citep{2009AIPC.1094..102C}.

WASP-150b is also relatively eccentric ($e$ = \eccentricityA). Using equation (3) from \citet{Adams2006} (and assuming $Q_p \sim 10^5$) we see that the approximate circularisation time-scale for this system is $\sim$ \circtimeA\,Gyr. Our analysis of this system gives the age of this star as \starageisoA\,Gyr, well below the circularisation time-scale, showing that the eccentricity is compatible with the other system parameters. On top of this it has been shown that more massive planets are more inclined to larger eccentricities \citep{2007A&A...464..779R} although their origin is still an open question. Additionally, we can use a simple scaling relation to predict the main-sequence lifetime of WASP-150. The following relation, $t_{MS} = t_{MS,\odot}\left(M/M_{\odot}\right)^{-2.5}$, where $t_{MS,\odot}$ is the main-sequence lifetime of the sun ($\sim$10\,Gyr), predicts the main-sequence lifetime of WASP-176 as $\sim$4.3\,Gyr, much longer than its estimated age.

For some examples of comparable high-mass, hot Jupiters in eccentric orbits see WASP-8b \citep[2.2\,$\rm M_J$, $e$ = 0.31;][]{Queloz2010}, Kepler-75 \citep[9.9\,$\rm M_J$, $e$ = 0.57;][]{Hebrard2013}, WASP-162b \citep[5.2\,$\rm M_J$,  $e$ = 0.43;][]{Hellier2018} and HAT-P-2b \citep[8.74\,$\rm M_J$, $e$ = 0.52;][]{2007ApJ...670..826B}. HAT-P-2b was the the first exoplanet discovered in this class and has been extensively studied in regards to planet-star interactions. The large mass and eccentricity of WASP-150b, coupled with its high eccentricity, make it an interesting addition to similar studies \citep{2013Icar..226.1642C,2016A&A...585L...2S,2017ApJ...836L..17D}. In particular the short period of this system allows for the collection of entire phase curves which will enable studies of the evolution of the planetary flux as a function of orbital phase \citep{2013AAS...22230202L}. Additionally, the eccentricity means that the planet cannot rotate synchronously. Thus the atmosphere should display time-dependent effects such as atmospheric radiative time constants and tidal luminosities \citep{2015arXiv150105685L}. Future studies may even be able to detect the manifestation of time-dependent storms \citep{2009Natur.457..562L}.

\subsection{WASP-176b}

WASP-176b is a slightly inflated hot Jupiter orbiting an \starclassspecB~host star on a \periodB\,day orbit. The MCMC best-fit results presented here give a planetary mass of \planetmassB\,$\rm M_J$ and a planetary radius of \planetradiusB\,$\rm R_J$ leading to a density of \planetdensityB\,$\rm \rho_J$. Using the same scaling relation as above predicts the main-sequence lifetime of WASP-176 as $\sim$4.77\,Gyr. Using the isochronal 
age estimated in section \ref{sec:Stellar parameters 176} indicates that the star has evolved past the end of its main-sequence life. This comparison, combined with the stellar surface gravity and density given in Table \ref{tab:MCMC parameters} indicates that the host is a sub-giant. This is further supported by its location on the HR diagram as seen in Figure \ref{fig:176_hr}. The planetary radius found here is approximately 25\% larger than expected for a coreless planet, when predicted using the planetary evolution models from \citet{Fortney2007}.

WASP-176b is similar to other hot Jupiters (WASP-54b: \citet{Faedi2013}; WASP-78b and WASP-79b: \citet{Smalley2012}; WASP142b: \citet{Hellier2016}; WASP-136b: \citet{Lam2016}) in that it receives stronger irradiation from its F-type host than it would from a G-type star, thus leading to inflation. Based on this increased radiation we suggest that the inflation mechanism behind WASP-176b may be due to the deposit of stellar irradiation into the planetary core, consistent with the Class I model presented in \citet{Lopez2016}. If so, then this planet may have experienced increasing inflation as its host moved off the main sequence. However this is only a possible solution, additional characterisation may give more information to this end.

WASP-176b is a potential target for atmospheric characterisation via transmission spectroscopy due to its low density and high equilibrium temperature. If we assume an atmosphere similar in composition to Jupiter ($\mu$ = 2.2u, where u = 1.66$\times 10^{-27}$\,kg is the atomic mass unit) then the scale height is found to be $H = k_BT/\mu g \sim$ 500\,km which is smaller, but comparable, to values seen in recent successful atmospheric retrievals \citep{Kirk2019}. However, with a magnitude of only $V$=\starmagVB, the follow-up would be a significant challenge.

\begin{acknowledgements}

We thank the anonymous referee for their helpful comments. BFC acknowledges a departmental scholarship from the University of Warwick. DP acknowledges support through a Merit Award from The Royal Society and from the Science and Technology Facilities Council (STFC) ST/P000495/1. MG and EJ are F.R.S.-FNRS Senior Research Associates. KB acknowledges support from the Erasmus+ programme. LD received support from the Gruber Foundation Fellowship. The research leading to these results has received funding from the European Research Council under the FP/2007-2013 ERC Grant Agreement number 336480 and from the ARC grant for Concerted Research Actions financed by the Wallonia-Brussels Federation. DLP acknowledges support from the Royal Society in the form of a Wolfson Merit award and STFC through the Warwick consolidated grant. This work was also partially supported by a grant from the Simons Foundation (PI Queloz, ID 327127). LM acknowledges support from the Italian Minister of Instruction, University and Research (MIUR) through FFABR 2017 fund. LM acknowledges support from the University of Rome Tor Vergata through ``Mission: Sustainability 2016'' fund. The Aristarchos telescope is operated on Helmos Observatory by the Institute for Astronomy, Astrophysics, Space Applications and Remote Sensing of the National Observatory of Athens. ACC acknowledges support from the Science and Technology Facilities Council (STFC) consolidated grant number ST/R000824/1. DJAB acknowledges support from the UK Space Agency (UKSA). TRAPPIST-South is a project funded by the Belgian Fonds (National) de la Recherche Scientifique (F.R.S.-FNRS) under grant FRFC 2.5.594.09.F. TRAPPIST-North is a project funded by the University of Li\`{e}ge, in collaboration with Cadi Ayyad University in Marrakech (Morocco). GD acknowledges support from CONICYT project Basal AFB-170002.

\end{acknowledgements}



\bibliography{WASP}{}

\begin{thebibliography}{}
\expandafter\ifx\csname natexlab\endcsname\relax\def\natexlab#1{#1}\fi
\providecommand{\url}[1]{\href{#1}{#1}}
\providecommand{\dodoi}[1]{doi:~\href{http://doi.org/#1}{\nolinkurl{#1}}}
\providecommand{\doeprint}[1]{\href{http://ascl.net/#1}{\nolinkurl{http://ascl.net/#1}}}
\providecommand{\doarXiv}[1]{\href{https://arxiv.org/abs/#1}{\nolinkurl{https://arxiv.org/abs/#1}}}

\bibitem[{Adams \& Laughlin(2006)}]{Adams2006}
Adams, F.~C., \& Laughlin, G. 2006,
  \href{https://iopscience.iop.org/article/10.1086/506145/pdf}{The
  Astrophysical Journal}, 649, 1004

\bibitem[{Bakos(2018)}]{Bakos2018}
Bakos, G.~{\' A}. 2018, Handbook of Exoplanets, 957–967,
  \dodoi{10.1007/978-3-319-55333-7\_111}

\bibitem[{{Bakos} {et~al.}(2007){Bakos}, {Kov{\'a}cs}, {Torres}, {Fischer},
  {Latham}, {Noyes}, {Sasselov}, {Mazeh}, {Shporer}, {Butler}, {Stefanik},
  {Fern{\'a}ndez}, {Sozzetti}, {P{\'a}l}, {Johnson}, {Marcy}, {Winn},
  {Sip{\H{o}}cz}, {L{\'a}z{\'a}r}, {Papp}, \& {S{\'a}ri}}]{2007ApJ...670..826B}
{Bakos}, G.~{\'A}., {Kov{\'a}cs}, G., {Torres}, G., {et~al.} 2007, \apj, 670,
  826, \dodoi{10.1086/521866}

\bibitem[{Baranne {et~al.}(1996)Baranne, Queloz, Mayor, {et~al.}}]{Baranne1996}
Baranne, A., Queloz, D., Mayor, M., {et~al.} 1996,
  \href{http://articles.adsabs.harvard.edu/cgi-bin/nph-iarticle_query?1996A\%26AS..119..373B&amp;data_type=PDF_HIGH&amp;whole_paper=YES&amp;type=PRINTER&amp;filetype=.pdf}{Astronomy
  and Astrophysics Supplement}, 119, 373

\bibitem[{Barbary(2016)}]{Barbary2016}
Barbary, K. 2016, \href{http://joss.theoj.org/papers/10.21105/joss.00058}{The
  Journal of Open Source Software}, 1

\bibitem[{Barkaoui {et~al.}(2019)Barkaoui, Burdanov, Hellier,
  {et~al.}}]{Barkaoui2018}
Barkaoui, K., Burdanov, A., Hellier, C., {et~al.} 2019,
  \href{https://iopscience.iop.org/article/10.3847/1538-3881/aaf422}{Astronomical
  Journal}, 157, 43

\bibitem[{Batalha(2014)}]{Batalha2014}
Batalha, N.~M. 2014,
  \href{http://www.pnas.org/content/pnas/111/35/12647.full.pdf}{Proceedings of
  the National Academy of Sciences}, 111, 12647

\bibitem[{Bertin \& Arnouts(1996)}]{Bertin1996}
Bertin, E., \& Arnouts, S. 1996,
  \href{http://articles.adsabs.harvard.edu/cgi-bin/nph-iarticle_query?1996A\%26AS..117..393B&amp;data_type=PDF_HIGH&amp;whole_paper=YES&amp;type=PRINTER&amp;filetype=.pdf}{Astronomy
  and Astrophysics Supplement}, 117, 393

\bibitem[{Boisse {et~al.}(2010)Boisse, Eggenberger, Santos,
  {et~al.}}]{Boisse2010}
Boisse, I., Eggenberger, A., Santos, N.~C., {et~al.} 2010,
  \href{https://www.aanda.org/articles/aa/pdf/2010/15/aa14909-10.pdf}{Astronomy
  and Astrophysics}, 523, A88

\bibitem[{Borucki {et~al.}(2010)Borucki, Koch, Basri, {et~al.}}]{Borucki2010}
Borucki, W.~J., Koch, D., Basri, G., {et~al.} 2010,
  \href{http://science.sciencemag.org/content/sci/327/5968/977.full.pdf}{Science},
  327, 977

\bibitem[{Bouchy {et~al.}(2009)Bouchy, Isambert, Lovis, {et~al.}}]{Bouchy2009b}
Bouchy, F., Isambert, J., Lovis, C., {et~al.} 2009, in
  \href{https://www.eas-journal.org/articles/eas/abs/2009/04/eas0937031/eas0937031.html}{EAS
  Publications Series}, ed. P.~{Kern}, Vol.~37, 247

\bibitem[{Boumis {et~al.}(2010)Boumis, Pollacco, Steele, {et~al.}}]{Boumis2010}
Boumis, P., Pollacco, D., Steele, I., {et~al.} 2010, in
  \href{http://articles.adsabs.harvard.edu/cgi-bin/nph-iarticle_query?2010ASPC..424..426B&amp;data_type=PDF_HIGH&amp;whole_paper=YES&amp;type=PRINTER&amp;filetype=.pdf}{Astronomical
  Society of the Pacific Conference Series}, Vol. 424, 9th International
  Conference of the Hellenic Astronomical Society, ed. K.~Tsinganos,
  D.~Hatzidimitriou, \& T.~Matsakos, 426

\bibitem[{Burdanov {et~al.}(2018)Burdanov, Delrez, Gillon, \&
  Jehin}]{Burdanov2018}
Burdanov, A., Delrez, L., Gillon, M., \& Jehin, E. 2018, Handbook of
  Exoplanets, 1007–1023, \dodoi{10.1007/978-3-319-55333-7_130}

\bibitem[{{C{\'e}bron} {et~al.}(2013){C{\'e}bron}, {Le Bars}, {Le Gal},
  {Moutou}, {Leconte}, \& {Sauret}}]{2013Icar..226.1642C}
{C{\'e}bron}, D., {Le Bars}, M., {Le Gal}, P., {et~al.} 2013, \icarus, 226,
  1642, \dodoi{10.1016/j.icarus.2012.12.017}

\bibitem[{{Chabrier} {et~al.}(2009){Chabrier}, {Baraffe}, {Leconte},
  {Gallardo}, \& {Barman}}]{2009AIPC.1094..102C}
{Chabrier}, G., {Baraffe}, I., {Leconte}, J., {Gallardo}, J., \& {Barman}, T.
  2009, in American Institute of Physics Conference Series, Vol. 1094, 15th
  Cambridge Workshop on Cool Stars, Stellar Systems, and the Sun, ed.
  E.~{Stempels}, 102--111, \dodoi{10.1063/1.3099078}

\bibitem[{Charbonneau {et~al.}(2000)Charbonneau, Brown, Latham,
  {et~al.}}]{Charbonneau2000}
Charbonneau, D., Brown, T.~M., Latham, D.~W., {et~al.} 2000,
  \href{http://iopscience.iop.org/article/10.1086/312457/pdf}{The Astrophysical
  Journal}, 529, L45

\bibitem[{Choi {et~al.}(2016)Choi, Dotter, Conroy, {et~al.}}]{Choi2016}
Choi, J., Dotter, A., Conroy, C., {et~al.} 2016,
  \href{https://iopscience.iop.org/article/10.3847/0004-637X/823/2/102/pdf}{The
  Astrophysical Journal}, 823, 102

\bibitem[{Ciceri {et~al.}(2013)Ciceri, Mancini, Southworth,
  {et~al.}}]{Ciceri2013}
Ciceri, S., Mancini, L., Southworth, J., {et~al.} 2013,
  \href{https://www.aanda.org/articles/aa/pdf/2013/09/aa21669-13.pdf}{Astronomy
  and Astrophysics}, 557, A30

\bibitem[{{Ciceri} {et~al.}(2013){Ciceri}, {Mancini}, {Southworth}, {Nikolov},
  {Bozza}, {Bruni}, {Calchi Novati}, {D'Ago}, \& {Henning}}]{ciceri13}
{Ciceri}, S., {Mancini}, L., {Southworth}, J., {et~al.} 2013,
  \href{https://ui.adsabs.harvard.edu/abs/2013A&A...557A..30C}{Astronomy and
  Astrophysics}, 557, A30, \dodoi{10.1051/0004-6361/201321669}

\bibitem[{Collier~Cameron {et~al.}(2007{\natexlab{a}})Collier~Cameron, Bouchy,
  H{\'e}brard, {et~al.}}]{CollierCameron2007b}
Collier~Cameron, A., Bouchy, F., H{\'e}brard, G., {et~al.} 2007{\natexlab{a}},
  \href{https://watermark.silverchair.com/mnras0375-0951.pdf?token=AQECAHi208BE49Ooan9kkhW_Ercy7Dm3ZL_9Cf3qfKAc485ysgAAAmswggJnBgkqhkiG9w0BBwagggJYMIICVAIBADCCAk0GCSqGSIb3DQEHATAeBglghkgBZQMEAS4wEQQMK1qtfhX8GCrqgMw6AgEQgIICHgLNNePBTpvN5rp6rXNQvjYN2AfDUBXHPGNBMFQWgiXkyJQlS3twJGN7dC2LcFpSrDpskdzWBtFF7LJr80zQWdDDM2kt2Gys4BWGHmD_466F5D62BQbo-fm-ujzSaCO7GThCNn3uy0g-o8a-m8xjhsGPyrpEptdxqFdQES7msbJVtpWt-YXabhTcbp0LuwCk4T_KEHGBOtMQNJh5dThz9FaeC4XE3DKsK7vOOI4kmkd7u9DhH_EEGAx0G8GfJb1fR2qy06l60sO-OWeD9Ntxay58f0WvJYvK_Zdm8n2eMXdgE10KfG-iUiGpqHgfFld6EGnragExAhsUHyGZmozPLAkH62c9Jkm733HCeXuhZt_W2dMewtABtZ_9rTPA5Vx3eGUV4ENIY5KGfBnzPU6d2kmG0-veQ9RgTP1yG-hsB0cTOsx22PmeuRPPYE9Ou6zHKSdvDdgXF1lHJt-eJUcbHjUpwy_mG6Z4T1hm0BtqFQ87_EN7Mbzcu3JTuFTKDmmRzzgBXzGN399SRqn3VaOOrmMkmuuDKgXAUXxxg82rnRRrwo1a5lptJITBc_fjwvRSayrapKnCcoo83WndY9sWIEgyZLBks2_m1RWql4vU7jXyt16QAd5EujJe1qgt7JfYizeWXirKeySp74I8PyG5xbVf8Dn-wXMySnJE6L8ZWFBkLtoxyC4fWhMx9so7NAY4QfYiHGF8PDV4Fth4sTKc}{Monthly
  Notices of the Royal Astronomical Society}, 375, 951

\bibitem[{Collier~Cameron {et~al.}(2006)Collier~Cameron, Pollacco, Street,
  {et~al.}}]{CollierCameron2006}
Collier~Cameron, A., Pollacco, D., Street, R.~A., {et~al.} 2006,
  \href{https://arxiv.org/pdf/astro-ph/0609418.pdf}{Monthly Notices of the
  Royal Astronomical Society}, 373, 799

\bibitem[{Collier~Cameron {et~al.}(2007{\natexlab{b}})Collier~Cameron, Wilson,
  West, {et~al.}}]{CollierCameron2007}
Collier~Cameron, A., Wilson, D.~M., West, R.~G., {et~al.} 2007{\natexlab{b}},
  \href{https://arxiv.org/pdf/0707.0417.pdf}{Monthly Notices of the Royal
  Astronomical Society}, 380, 1230

\bibitem[{Craig {et~al.}(2015)Craig, Crawford, Deil, {et~al.}}]{Craig2015}
Craig, M.~W., Crawford, S.~M., Deil, C., {et~al.} 2015,
  \href{http://adsabs.harvard.edu/abs/2015ascl.soft10007C}{Astrophysics Source
  Code Library}, 1510.007, 393

\bibitem[{Cumming {et~al.}(2008)Cumming, Butler, Marcy, {et~al.}}]{Cumming2008}
Cumming, A., Butler, R.~P., Marcy, G.~W., {et~al.} 2008,
  \href{http://iopscience.iop.org/article/10.1086/588487/pdf}{Publications of
  the Astronomical Society of the Pacific}, 120, 531

\bibitem[{{de Wit} {et~al.}(2017){de Wit}, {Lewis}, {Knutson}, {Fuller},
  {Antoci}, {Fulton}, {Laughlin}, {Deming}, {Shporer}, {Batygin}, {Cowan},
  {Agol}, {Burrows}, {Fortney}, {Langton}, \& {Showman}}]{2017ApJ...836L..17D}
{de Wit}, J., {Lewis}, N.~K., {Knutson}, H.~A., {et~al.} 2017, \apjl, 836, L17,
  \dodoi{10.3847/2041-8213/836/2/L17}

\bibitem[{Delrez {et~al.}(2018)Delrez, Gillon, Queloz, {et~al.}}]{Delrez2018}
Delrez, L., Gillon, M., Queloz, D., {et~al.} 2018,
  \href{https://arxiv.org/pdf/1806.11205.pdf}{ArXiv e-prints}, 1806.11205

\bibitem[{Dotter(2016)}]{Dotter2016}
Dotter, A. 2016,
  \href{https://iopscience.iop.org/article/10.3847/0067-0049/222/1/8/pdf}{The
  Astrophysical Journal Supplement Series}, 222, 8

\bibitem[{Doyle {et~al.}(2013)Doyle, Smalley, Maxted, {et~al.}}]{Doyle2013}
Doyle, A.~P., Smalley, B., Maxted, P. F.~L., {et~al.} 2013,
  \href{https://arxiv.org/pdf/1210.5931.pdf}{Monthly Notices of the Royal
  Astronomical Society}, 428, 3164

\bibitem[{Eastman(2017)}]{Eastman2017}
Eastman, J. 2017, \href{http://ascl.net/1710.003}{Astrophysics Source Code
  Library}

\bibitem[{Eastman {et~al.}(2019)Eastman, Rodriguez, Agol,
  {et~al.}}]{Eastman2019}
Eastman, J.~D., Rodriguez, J.~E., Agol, E., {et~al.} 2019,
  \href{https://arxiv.org/pdf/1907.09480.pdf}{arXiv e-prints}, 1907.09480

\bibitem[{Faedi {et~al.}(2013)Faedi, Pollacco, Barros, {et~al.}}]{Faedi2013}
Faedi, F., Pollacco, D., Barros, S. C.~C., {et~al.} 2013,
  \href{https://arxiv.org/pdf/1210.2329.pdf}{Astronomy and Astrophysics}, 551,
  A73

\bibitem[{{Feinstein} {et~al.}(2019){Feinstein}, {Montet}, {Foreman-Mackey},
  {Bedell}, {Saunders}, {Bean}, {Christiansen}, {Hedges}, {Luger}, {Scolnic},
  \& {Cardoso}}]{2019PASP..131i4502F}
{Feinstein}, A.~D., {Montet}, B.~T., {Foreman-Mackey}, D., {et~al.} 2019,
  \pasp, 131, 094502, \dodoi{10.1088/1538-3873/ab291c}

\bibitem[{Ford(2006)}]{Ford2006}
Ford, E.~B. 2006,
  \href{https://iopscience.iop.org/article/10.1086/500802/pdf}{The
  Astrophysical Journal}, 642, 505

\bibitem[{Fortney {et~al.}(2007)Fortney, Marley, \& Barnes}]{Fortney2007}
Fortney, J.~J., Marley, M.~S., \& Barnes, J.~W. 2007,
  \href{https://arxiv.org/pdf/astro-ph/0612671.pdf}{The Astronomical Journal},
  659, 1661

\bibitem[{{Gaia Collaboration} {et~al.}(2018){Gaia Collaboration}, Brown,
  Vallenari, {et~al.}}]{Gaia2018}
{Gaia Collaboration}, Brown, A. G.~A., Vallenari, A., {et~al.} 2018,
  \href{https://arxiv.org/pdf/1804.09365.pdf}{ArXiv e-prints}, 1804.09365

\bibitem[{{Gaia Collaboration} {et~al.}(2016){Gaia Collaboration}, Prusti,
  de~Bruijne, {et~al.}}]{Gaia2016}
{Gaia Collaboration}, Prusti, T., de~Bruijne, J. H.~J., {et~al.} 2016,
  \href{https://arxiv.org/pdf/1609.04153.pdf}{Astronomy and Astrophysics}, 595,
  A1

\bibitem[{Gardner {et~al.}(2006)Gardner, Mather, Clampin,
  {et~al.}}]{Gardner2006}
Gardner, J.~P., Mather, J.~C., Clampin, M., {et~al.} 2006,
  \href{https://arxiv.org/pdf/astro-ph/0606175.pdf}{Space Science Reviews},
  123, 485

\bibitem[{{Gauza} {et~al.}(2015){Gauza}, {B{\'e}jar}, {Rebolo}, {{\'A}lvarez},
  {Bihain}, {Zapatero Osorio}, {Caballero}, {Telesco}, \& {Packham}}]{gauza15}
{Gauza}, B., {B{\'e}jar}, V.~J.~S., {Rebolo}, R., {et~al.} 2015,
  \href{https://ui.adsabs.harvard.edu/abs/2015MNRAS.452.1677G}{Monthly Notices
  of the Royal Astronomical Society}, 452, 1677, \dodoi{10.1093/mnras/stv1350}

\bibitem[{Gelman {et~al.}(2003)Gelman, Carlin, Stern, {et~al.}}]{Gelman2003}
Gelman, A., Carlin, J.~B., Stern, H.~S., {et~al.} 2003, Bayesian Data Analysis,
  2nd edn. (Chapman \& Hall)

\bibitem[{{Gibson} {et~al.}(2008){Gibson}, {Pollacco}, {Simpson}, {Joshi},
  {Todd}, {Benn}, {Christian}, {Hrudkov{\'a}}, {Keenan}, {Meaburn}, {Skillen},
  \& {Steele}}]{gibson08}
{Gibson}, N.~P., {Pollacco}, D., {Simpson}, E.~K., {et~al.} 2008,
  \href{https://ui.adsabs.harvard.edu/abs/2008A&A...492..603G}{Astronomy and
  Astrophysics}, 492, 603, \dodoi{10.1051/0004-6361:200811015}

\bibitem[{Gillon(2018)}]{Gillon2018}
Gillon, M. 2018,
  \href{http://adsabs.harvard.edu/abs/2018NatAs...2..344G}{Nature Astronomy},
  2, 344

\bibitem[{Gillon {et~al.}(2013)Gillon, Anderson, Collier-Cameron,
  {et~al.}}]{Gillon2013}
Gillon, M., Anderson, D.~R., Collier-Cameron, A., {et~al.} 2013,
  \href{https://www.aanda.org/articles/aa/pdf/2013/04/aa20561-12.pdf}{Astronomy
  and Astrophysics}, 552, A82

\bibitem[{Gillon {et~al.}(2011)Gillon, Jehin, Magain, {et~al.}}]{Gillon2011}
Gillon, M., Jehin, E., Magain, P., {et~al.} 2011,
  \href{https://arxiv.org/pdf/1101.5807.pdf}{{EPJ} Web of Conferences}, 11,
  06002

\bibitem[{Gillon {et~al.}(2017)Gillon, Triaud, Demory, {et~al.}}]{Gillon2017}
Gillon, M., Triaud, A. H. M.~J., Demory, B.-O., {et~al.} 2017,
  \href{https://arxiv.org/ftp/arxiv/papers/1703/1703.01424.pdf}{nature}, 542,
  456

\bibitem[{Grubbs(1950)}]{grubbs1950}
Grubbs, F.~E. 1950, Ann. Math. Statist., 21, 27,
  \dodoi{10.1214/aoms/1177729885}

\bibitem[{H{\'e}brard {et~al.}(2013)H{\'e}brard, Almenara, Santerne,
  {et~al.}}]{Hebrard2013}
H{\'e}brard, G., Almenara, J.~M., Santerne, A., {et~al.} 2013,
  \href{https://www.aanda.org/articles/aa/pdf/2013/06/aa21394-13.pdf}{Astronomy
  and Astrophysics}, 554, A114

\bibitem[{Hellier {et~al.}(2018)Hellier, Anderson, Bouchy,
  {et~al.}}]{Hellier2018}
Hellier, C., Anderson, D.~R., Bouchy, F., {et~al.} 2018,
  \href{https://watermark.silverchair.com/sty2741.pdf?token=AQECAHi208BE49Ooan9kkhW_Ercy7Dm3ZL_9Cf3qfKAc485ysgAAAlgwggJUBgkqhkiG9w0BBwagggJFMIICQQIBADCCAjoGCSqGSIb3DQEHATAeBglghkgBZQMEAS4wEQQMFwWe50bTQ_VCPOmnAgEQgIICC7v8CJ1pdxoMYawbm8qqbmAugZR4kQNiFDPXiBcuEJXW1tr94txsrxAlZrdT4bUJDhiGJjLXa5UaADypzCokw8x4lcBQ3AST914Juw2QU7iSLtb9WyydJLtFcfPrmKkWYYGfoKA7Bqso_i6IpSHFtUOv7Cu33IoQ3-vgjgn-hrkcuZlXpw3OnsYkZDiZS-fD9Obrcz5ZmZJxDiWzWL7nmVGT_95Y0xPZYaBP-rNpN5x7n38AHkkPrc_MU0AhrjbYAOuR1QcC1uoblbGMuhiFjMvu47tK0yS2rKlSQmd8cACJlWHwK4yFHNlAxltuOLzdQ0Rzme-kAJWwdah4NHZKE0w7ZTY_g0J_r_ndU88Nf3zh26H-wwPER4lM9iKEe0qIXU9hHTfoE0DfnyEly6V7SFiVDdI6zII4Xr7TXXZHqOJ4VgVA7JMQr_UCY-19Olk37fS8TdB0WCtgqvTHfQ6Lkp5QU_2Pce5jpX6EDqocIfqtU18j0AUarS1HtzL1zRzRHxlNn8gldQE4wDp6-r0aAN_IM4AP17MVUpDVIVRWwghlqnF3sdEi45ISg4JLLnLaUb5ZBjNbsl2_Nl9jzeXAxIFPiZjE8r40ut0OZmd9tV-7BNAtZxxaenj0NstLsbXdF0jxmtgWeFwg1L-TyCVfTHF7s5RMAmuiY3Q7mTRrHZnLGSNg4VPaEiDZvLg}{Monthly
  Notices of the Royal Astronomical Society}, 482, 1379

\bibitem[{Hellier {et~al.}(2016)Hellier, Anderson, Cameron,
  {et~al.}}]{Hellier2016}
Hellier, C., Anderson, D.~R., Cameron, A.~C., {et~al.} 2016,
  \href{https://arxiv.org/pdf/1604.04195.pdf}{Monthly Notices of the Royal
  Astronomical Society}, 465, 3693

\bibitem[{Henry {et~al.}(2000)Henry, Marcy, Butler, {et~al.}}]{Henry2000}
Henry, G.~W., Marcy, G.~W., Butler, R.~P., {et~al.} 2000,
  \href{http://iopscience.iop.org/article/10.1086/312458/pdf}{The Astrophysical
  Journal}, 529, L41

\bibitem[{Howell {et~al.}(2014)Howell, Sobeck, Haas, {et~al.}}]{Howell2014}
Howell, S.~B., Sobeck, C., Haas, M., {et~al.} 2014,
  \href{http://iopscience.iop.org/article/10.1086/676406/pdf}{Publications of
  the Astronomical Society of the Pacific}, 126, 398

\bibitem[{Jehin {et~al.}(2011)Jehin, Gillon, Queloz, {et~al.}}]{Jehin2011}
Jehin, E., Gillon, M., Queloz, D., {et~al.} 2011,
  \href{http://www.eso.org/sci/publications/messenger/archive/no.145-sep11/messenger-no145-2-6.pdf}{The
  Messenger}, 145, 2

\bibitem[{{Jehin} {et~al.}(2018){Jehin}, {Gillon}, {Queloz}, {Delrez},
  {Burdanov}, {Murray}, {Sohy}, {Ducrot}, {Sebastian}, {Thompson}, {McCormac},
  {Almleaky}, {Burgasser}, {Demory}, {de Wit}, {Barkaoui}, {Pozuelos},
  {Triaud}, \& {Grootel}}]{Jehin2018}
{Jehin}, E., {Gillon}, M., {Queloz}, D., {et~al.} 2018, The Messenger, 174, 2,
  \dodoi{10.18727/0722-6691/5105}

\bibitem[{{J{\'o}dar} {et~al.}(2013){J{\'o}dar}, {P{\'e}rez-Garrido},
  {D{\'{\i}}az-S{\'a}nchez}, {Vill{\'o}}, {Rebolo}, \&
  {P{\'e}rez-Prieto}}]{jodar13}
{J{\'o}dar}, E., {P{\'e}rez-Garrido}, A., {D{\'{\i}}az-S{\'a}nchez}, A.,
  {et~al.} 2013,
  \href{http://adsabs.harvard.edu/abs/2013MNRAS.429..859J}{Monthly Notices of
  the Royal Astronomical Society}, 429, 859, \dodoi{10.1093/mnras/sts382}

\bibitem[{{Kirk} {et~al.}(2019){Kirk}, {L{\'o}pez-Morales}, {Wheatley},
  {Weaver}, {Skillen}, {Louden}, {McCormac}, \& {Espinoza}}]{Kirk2019}
{Kirk}, J., {L{\'o}pez-Morales}, M., {Wheatley}, P.~J., {et~al.} 2019, \aj,
  158, 144, \dodoi{10.3847/1538-3881/ab397d}

\bibitem[{Kov{\'a}cs {et~al.}(2002)Kov{\'a}cs, Zucker, \& Mazeh}]{Kovacs2002}
Kov{\'a}cs, G., Zucker, S., \& Mazeh, T. 2002,
  \href{https://www.aanda.org/articles/aa/pdf/2002/31/aa2422.pdf}{Astronomy and
  Astrophysics}, 391, 369

\bibitem[{{Labadie} {et~al.}(2010){Labadie}, {Rebolo}, {Femen{\'{\i}}a},
  {Vill{\'o}}, {D{\'{\i}}az-S{\'a}nchez}, {Oscoz}, {L{\'o}pez},
  {P{\'e}rez-Prieto}, {P{\'e}rez-Garrido}, {Hildebrandt},
  {B{\'e}jar-S{\'a}nchez}, {Jos{\'e} Piqueras}, \&
  {Rodr{\'{\i}}guez}}]{labadie10}
{Labadie}, L., {Rebolo}, R., {Femen{\'{\i}}a}, B., {et~al.} 2010, in
  \href{http://adsabs.harvard.edu/abs/2010SPIE.7735E..0XL}{Society of
  Photo-Optical Instrumentation Engineers (SPIE) Conference Series}, Vol. 7735,
  Ground-based and Airborne Instrumentation for Astronomy III, 77350X,
  \dodoi{10.1117/12.857998}

\bibitem[{Lam {et~al.}(2016)Lam, Faedi, Brown, {et~al.}}]{Lam2016}
Lam, K. W.~F., Faedi, F., Brown, D. J.~A., {et~al.} 2016,
  \href{https://arxiv.org/pdf/1607.07859.pdf}{Astronomy and Astrophysics}, 599,
  A3

\bibitem[{{Laughlin} {et~al.}(2009){Laughlin}, {Deming}, {Langton}, {Kasen},
  {Vogt}, {Butler}, {Rivera}, \& {Meschiari}}]{2009Natur.457..562L}
{Laughlin}, G., {Deming}, D., {Langton}, J., {et~al.} 2009, \nat, 457, 562,
  \dodoi{10.1038/nature07649}

\bibitem[{{Laughlin} \& {Lissauer}(2015)}]{2015arXiv150105685L}
{Laughlin}, G., \& {Lissauer}, J.~J. 2015, arXiv e-prints, arXiv:1501.05685.
\newblock \doarXiv{1501.05685}

\bibitem[{{Lewis} {et~al.}(2013){Lewis}, {Showman}, {Fortney}, {Knutson}, \&
  {Marley}}]{2013AAS...22230202L}
{Lewis}, N., {Showman}, A.~P., {Fortney}, J.~J., {Knutson}, H., \& {Marley},
  M.~S. 2013, in American Astronomical Society Meeting Abstracts, Vol. 222,
  American Astronomical Society Meeting Abstracts, 302.02

\bibitem[{Lopez \& Fortney(2016)}]{Lopez2016}
Lopez, E.~D., \& Fortney, J.~J. 2016,
  \href{https://arxiv.org/pdf/1510.00067.pdf}{The Astrophysical Journal}, 818,
  4

\bibitem[{Maxted {et~al.}(2015)Maxted, Serenelli, \& Southworth}]{Maxted2015a}
Maxted, P. F.~L., Serenelli, A.~M., \& Southworth, J. 2015,
  \href{https://www.aanda.org/articles/aa/pdf/2015/03/aa25331-14.pdf}{Astronomy
  and Astrophysics}, 575, A36

\bibitem[{Mayor {et~al.}(2011)Mayor, Marmier, Lovis, {et~al.}}]{Mayor2011}
Mayor, M., Marmier, M., Lovis, C., {et~al.} 2011,
  \href{https://arxiv.org/pdf/1109.2497.pdf}{ArXiv e-prints}, 1109.2497

\bibitem[{McCormac {et~al.}(2014)McCormac, Skillen, Pollacco,
  {et~al.}}]{McCormac2014}
McCormac, J., Skillen, I., Pollacco, D., {et~al.} 2014,
  \href{https://arxiv.org/pdf/1312.5880.pdf}{Monthly Notices of the Royal
  Astronomical Society}, 438, 3383

\bibitem[{{Murga} {et~al.}(2014){Murga}, {Oscoz}, {L{\'o}pez}, {Campo},
  {Etxegarai}, \& {Pall{\'e}}}]{murga14}
{Murga}, G., {Oscoz}, A., {L{\'o}pez}, R., {et~al.} 2014, in
  \href{https://ui.adsabs.harvard.edu/abs/2014SPIE.9147E..6QM}{Society of
  Photo-Optical Instrumentation Engineers (SPIE) Conference Series}, Vol. 9147,
  \procspie, 91476Q, \dodoi{10.1117/12.2057127}

\bibitem[{{Oscoz} {et~al.}(2008){Oscoz}, {Rebolo}, {L{\'o}pez},
  {P{\'e}rez-Garrido}, {P{\'e}rez}, {Hildebrandt}, {Rodr{\'{\i}}guez},
  {Piqueras}, {Vill{\'o}}, {Gonz{\'a}lez}, {Barrena}, {G{\'o}mez},
  {Garc{\'{\i}}a-Hern{\'a}ndez}, {Monta{\~n}{\'e}s}, {Rosenberg}, {Cadavid},
  {Calcines}, {D{\'{\i}}az-S{\'a}nchez}, {Kohley}, {Mart{\'{\i}}n},
  {Pe{\~n}ate}, \& {S{\'a}nchez}}]{oscoz08}
{Oscoz}, A., {Rebolo}, R., {L{\'o}pez}, R., {et~al.} 2008, in
  \href{http://adsabs.harvard.edu/abs/2008SPIE.7014E..47O}{Society of
  Photo-Optical Instrumentation Engineers (SPIE) Conference Series}, Vol. 7014,
  Ground-based and Airborne Instrumentation for Astronomy II, 701447,
  \dodoi{10.1117/12.788834}

\bibitem[{{Pascale} {et~al.}(2018){Pascale}, {Bezawada}, {Barstow}, {Beaulieu},
  {Bowles}, {Coud{\'e} du Foresto}, {Coustenis}, {Decin}, {Drossart},
  {Eccleston}, {Encrenaz}, {Forget}, {Griffin}, {G{\"u}del}, {Hartogh},
  {Heske}, {Lagage}, {Leconte}, {Malaguti}, {Micela}, {Middleton}, {Min},
  {Moneti}, {Morales}, {Mugnai}, {Ollivier}, {Pace}, {Papageorgiou},
  {Pilbratt}, {Puig}, {Rataj}, {Ray}, {Ribas}, {Rocchetto}, {Sarkar}, {Selsis},
  {Taylor}, {Tennyson}, {Tinetti}, {Turrini}, {Vandenbussche}, {Venot},
  {Waldmann}, {Wolkenberg}, {Wright}, {Zapatero Osorio}, \&
  {Zingales}}]{Pascale2018}
{Pascale}, E., {Bezawada}, N., {Barstow}, J., {et~al.} 2018, in Society of
  Photo-Optical Instrumentation Engineers (SPIE) Conference Series, Vol. 10698,
  \procspie, 106980H, \dodoi{10.1117/12.2311838}

\bibitem[{Pepe {et~al.}(2017)Pepe, Bouchy, Mayor, {et~al.}}]{Pepe2017}
Pepe, F., Bouchy, F., Mayor, M., {et~al.} 2017, Handbook of Exoplanets, Edited
  by Hans J. Deeg and Juan Antonio Belmonte
  (\href{https://link.springer.com/referenceworkentry/10.1007\%2F978-3-319-30648-3_190-1}{Springer
  Reference Works}), 190

\bibitem[{Pepe {et~al.}(2002)Pepe, Mayor, Galland, {et~al.}}]{Pepe2002}
Pepe, F., Mayor, M., Galland, F., {et~al.} 2002,
  \href{https://www.aanda.org/articles/aa/pdf/2002/23/aah3477.pdf}{Astronomy
  and Astrophysics}, 388, 632

\bibitem[{Pepper {et~al.}(2007)Pepper, Pogge, DePoy, {et~al.}}]{Pepper2007}
Pepper, J., Pogge, R.~W., DePoy, D.~L., {et~al.} 2007,
  \href{http://iopscience.iop.org/article/10.1086/521836/pdf}{Publications of
  the Astronomical Society of the Pacific}, 119, 923

\bibitem[{Perruchot {et~al.}(2008)Perruchot, Kohler, Bouchy,
  {et~al.}}]{Perruchot2008}
Perruchot, S., Kohler, D., Bouchy, F., {et~al.} 2008, in
  \href{https://www.spiedigitallibrary.org/conference-proceedings-of-spie/7014/1/The-SOPHIE-spectrograph--design-and-technical-key-points-for/10.1117/12.787379.short?SSO=1}{Society
  of Photo-Optical Instrumentation Engineers (SPIE) Conference Series}, Vol.
  7014, Ground-based and Airborne Instrumentation for Astronomy II, 70140J

\bibitem[{Pollacco {et~al.}(2006)Pollacco, Skillen, {et~al.}}]{Pollacco2006}
Pollacco, D.~L., Skillen, I. Collier~Cameron, A., {et~al.} 2006,
  \href{https://arxiv.org/pdf/astro-ph/0608454.pdf}{Publications of the
  Astronomical Society of the Pacific}, 118, 1407

\bibitem[{Queloz {et~al.}(2010)Queloz, Anderson, Collier~Cameron,
  {et~al.}}]{Queloz2010}
Queloz, D., Anderson, D.~R., Collier~Cameron, A., {et~al.} 2010,
  \href{https://www.aanda.org/articles/aa/pdf/2010/09/aa14768-10.pdf}{Astronomy
  and Astrophysics}, 517, L1

\bibitem[{Queloz {et~al.}(2001)Queloz, Henry, Sivan, {et~al.}}]{Queloz2001}
Queloz, D., Henry, G.~W., Sivan, J.~P., {et~al.} 2001,
  \href{https://www.aanda.org/articles/aa/pdf/2001/43/aa1802.pdf}{Astronomy and
  Astrophysics}, 379, 279

\bibitem[{Queloz {et~al.}(2000)Queloz, Mayor, Weber, {et~al.}}]{Queloz2000}
Queloz, D., Mayor, M., Weber, L., {et~al.} 2000,
  \href{http://articles.adsabs.harvard.edu/cgi-bin/nph-iarticle_query?2000A\%26A...354...99Q&amp;data_type=PDF_HIGH&amp;whole_paper=YES&amp;type=PRINTER&amp;filetype=.pdf}{Astronomy
  and Astrophysics}, 354, 99

\bibitem[{{Rauer} {et~al.}(2016){Rauer}, {Aerts}, {Cabrera}, \& {PLATO
  Team}}]{Rauer2016}
{Rauer}, H., {Aerts}, C., {Cabrera}, J., \& {PLATO Team}. 2016, Astronomische
  Nachrichten, 337, 961, \dodoi{10.1002/asna.201612408}

\bibitem[{{Ribas} \& {Miralda-Escud{\'e}}(2007)}]{2007A&A...464..779R}
{Ribas}, I., \& {Miralda-Escud{\'e}}, J. 2007, \aap, 464, 779,
  \dodoi{10.1051/0004-6361:20065726}

\bibitem[{Ricker {et~al.}(2015)Ricker, Winn, Vanderspek, {et~al.}}]{Ricker2015}
Ricker, G.~R., Winn, J.~N., Vanderspek, R., {et~al.} 2015,
  \href{https://arxiv.org/pdf/1406.0151.pdf}{Journal of Astronomical
  Telescopes, Instruments and Systems}, 1, 014003

\bibitem[{{Ricker} {et~al.}(2015){Ricker}, {Winn}, {Vanderspek}, {Latham},
  {Bakos}, {Bean}, {Berta-Thompson}, {Brown}, {Buchhave}, {Butler}, {Butler},
  {Chaplin}, {Charbonneau}, {Christensen-Dalsgaard}, {Clampin}, {Deming},
  {Doty}, {De Lee}, {Dressing}, {Dunham}, {Endl}, {Fressin}, {Ge}, {Henning},
  {Holman}, {Howard}, {Ida}, {Jenkins}, {Jernigan}, {Johnson}, {Kaltenegger},
  {Kawai}, {Kjeldsen}, {Laughlin}, {Levine}, {Lin}, {Lissauer}, {MacQueen},
  {Marcy}, {McCullough}, {Morton}, {Narita}, {Paegert}, {Palle}, {Pepe},
  {Pepper}, {Quirrenbach}, {Rinehart}, {Sasselov}, {Sato}, {Seager},
  {Sozzetti}, {Stassun}, {Sullivan}, {Szentgyorgyi}, {Torres}, {Udry}, \&
  {Villasenor}}]{2015JATIS...1a4003R}
{Ricker}, G.~R., {Winn}, J.~N., {Vanderspek}, R., {et~al.} 2015, Journal of
  Astronomical Telescopes, Instruments, and Systems, 1, 014003,
  \dodoi{10.1117/1.JATIS.1.1.014003}

\bibitem[{{Salz} {et~al.}(2016){Salz}, {Schneider}, {Czesla}, \&
  {Schmitt}}]{2016A&A...585L...2S}
{Salz}, M., {Schneider}, P.~C., {Czesla}, S., \& {Schmitt}, J.~H.~M.~M. 2016,
  \aap, 585, L2, \dodoi{10.1051/0004-6361/201527042}

\bibitem[{Schanche {et~al.}(2019)Schanche, Collier~Cameron, H{\'e}brard,
  {et~al.}}]{Schanche2019}
Schanche, N., Collier~Cameron, A., H{\'e}brard, G., {et~al.} 2019,
  \href{https://watermark.silverchair.com/sty3146.pdf?token=AQECAHi208BE49Ooan9kkhW_Ercy7Dm3ZL_9Cf3qfKAc485ysgAAAlkwggJVBgkqhkiG9w0BBwagggJGMIICQgIBADCCAjsGCSqGSIb3DQEHATAeBglghkgBZQMEAS4wEQQM6bD_aaYK2bWdY-uJAgEQgIICDJVS618I0zwbIRUkMmPd89Nfx6j700LPsYEUj7syrG_-b2b3OSMVUdXQGyIsjM78I2J4CVnw8IW8JDywdjpo56fLPJ03eijmGH61Y5cKry3BzeY4hqbX1kZFoDFVJdakx_Cw5q21owIMwSFuF4C2aIQaQFVvBB6DCGPNEc_8ccQtdWsGE4Pj6GiBz09imH9mHibytUssfYvQIFVie38_4AcQ3PEyHDewi3LY2qPMP-VYg9nomcdBCHo_rQHcO0GHy3FI-NhpmKo-_XrwNIFSjdyWHAPr_in0W_0VioCInc75giUqbpyXpeOWU5_mLqPCPLgG1pVLObg1qS4n2uj6cUV4nTkEJDNiuNcUwc04lHCR3HFTeKDJzNAKCFPtHOZCQtARcYv5hcUivJS_jjkA7rGr8b2E1oP7KhpC7gD1Y3B-azGf9-c0EGU2_QLNJ_obxfEJ_CVXot505ePbnWsHxPwFfCkG1GrB3YpeFSVIgusPxU8jUBhacfPed7hYh-aCx7irMAnSZz_hVRorqh0zMYcqPigvoaj0BSSNlayEf0cn0wNKzsm2dhhkO-QEZF8X5yM8dPyHWVvekbi-ZuU77fZDHTzpLAFvcZ99-ao1TL2GpnVCbvfMYT2St635OtCkvapye7CztZ_umh1THNde_nd1Ye5Hde8Nvz9IgOx-jp6pH9x6M0EPEkVLNPeX}{Monthly
  Notices of the Royal Astronomical Society}, 483, 5534

\bibitem[{Schlafly \& Finkbeiner(2011)}]{Schlafly2011}
Schlafly, E.~F., \& Finkbeiner, D.~P. 2011,
  \href{https://iopscience.iop.org/article/10.1088/0004-637X/737/2/103/pdf}{The
  Astrophysical Journal}, 737, 103

\bibitem[{Schlegel {et~al.}(1998)Schlegel, Finkbeiner, \& Davis}]{Schlegel1998}
Schlegel, D.~J., Finkbeiner, D.~P., \& Davis, M. 1998,
  \href{https://iopscience.iop.org/article/10.1086/305772/pdf}{The
  Astrophysical Journal}, 500, 525

\bibitem[{Smalley {et~al.}(2012)Smalley, Anderson, Collier-Cameron,
  {et~al.}}]{Smalley2012}
Smalley, B., Anderson, D.~R., Collier-Cameron, A., {et~al.} 2012,
  \href{https://arxiv.org/pdf/1206.1177.pdf}{Astronomy and Astrophysics}, 547,
  A61

\bibitem[{Southworth {et~al.}(2014)Southworth, Hinse, Burgdorf,
  {et~al.}}]{Southworth2014}
Southworth, J., Hinse, T.~C., Burgdorf, M., {et~al.} 2014,
  \href{https://arxiv.org/pdf/1407.6253.pdf}{Monthly Notices of the Royal
  Astronomical Society}, 444, 776

\bibitem[{Southworth {et~al.}(2009)Southworth, Hinse, J{\o}rgensen,
  {et~al.}}]{Southworth2009}
Southworth, J., Hinse, T.~C., J{\o}rgensen, U.~G., {et~al.} 2009,
  \href{https://arxiv.org/pdf/0903.2139.pdf}{Monthly Notices of the Royal
  Astronomical Society}, 396, 1023

\bibitem[{{Steele} {et~al.}(2008){Steele}, {Bates}, {Gibson}, {Keenan},
  {Meaburn}, {Mottram}, {Pollacco}, \& {Todd}}]{steele08}
{Steele}, I.~A., {Bates}, S.~D., {Gibson}, N., {et~al.} 2008, in
  \href{https://ui.adsabs.harvard.edu/abs/2008SPIE.7014E..6JS}{Society of
  Photo-Optical Instrumentation Engineers (SPIE) Conference Series}, Vol. 7014,
  \procspie, 70146J, \dodoi{10.1117/12.787889}

\bibitem[{{Stetson}(1987)}]{stetson87}
{Stetson}, P.~B. 1987,
  \href{https://ui.adsabs.harvard.edu/abs/1987PASP...99..191S}{Publications of
  the Astronomical Society of the Pacific}, 99, 191, \dodoi{10.1086/131977}

\bibitem[{Tamuz {et~al.}(2005)Tamuz, Mazeh, \& Zucker}]{Tamuz2005}
Tamuz, O., Mazeh, T., \& Zucker, S. 2005,
  \href{https://arxiv.org/pdf/astro-ph/0502056.pdf}{Monthly Notices of the
  Royal Astronomical Society}, 356, 1466

\bibitem[{{Tody}(1986)}]{tody86}
{Tody}, D. 1986, in
  \href{https://ui.adsabs.harvard.edu/abs/1986SPIE..627..733T}{Society of
  Photo-Optical Instrumentation Engineers (SPIE) Conference Series}, Vol. 627,
  \procspie, ed. D.~L. {Crawford}, 733, \dodoi{10.1117/12.968154}

\bibitem[{Tody(1986)}]{Tody1986}
Tody, D. 1986, in Proceedings of the Meeting, Tucson, AZ, March 4-8, 1986, Vol.
  627, Instrumentation in astronomy VI, ed. D.~L. Crawford (Bellingham, WA:
  \href{https://www.spiedigitallibrary.org/conference-proceedings-of-spie/0627/1/The-Iraf-Data-Reduction-And-Analysis-System/10.1117/12.968154.short?SSO=1}{Society
  of Photo-Optical Instrumentation Engineers (SPIE) Conference Series}), 733

\bibitem[{{Tody}(1993)}]{tody93}
{Tody}, D. 1993, in
  \href{https://ui.adsabs.harvard.edu/abs/1993ASPC...52..173T}{Astronomical
  Society of the Pacific Conference Series}, Vol.~52, Astronomical Data
  Analysis Software and Systems II, ed. R.~J. {Hanisch}, R.~J.~V. {Brissenden},
  \& J.~{Barnes}, 173

\bibitem[{Weiss \& Schlattl(2008)}]{Weiss2008}
Weiss, A., \& Schlattl. 2008,
  \href{https://link.springer.com/content/pdf/10.1007\%2Fs10509-007-9606-5.pdf}{Astrophysics
  and Space Science}, 316, 99

\bibitem[{Wheatley {et~al.}(2017)Wheatley, West, Goad, {et~al.}}]{Wheatley2017}
Wheatley, P.~J., West, R.~G., Goad, M.~R., {et~al.} 2017,
  \href{https://arxiv.org/pdf/1710.11100.pdf}{Monthly Notices of the Royal
  Astronomical Society}

\bibitem[{{Winn}(2010)}]{2010exop.book...55W}
{Winn}, J.~N. 2010, {Exoplanet Transits and Occultations} (University of
  Arizona Press), 55--77

\end{thebibliography}
\bibliographystyle{aasjournal}

\end{document}